\documentclass[onecolumn,notitlepage]{revtex4-1}
\usepackage{relsize}
\usepackage{upgreek}

\usepackage{amsmath,amssymb,graphicx,hyperref}
\usepackage{array,booktabs}
\usepackage{matlab-prettifier}
\usepackage{listings}
\usepackage{bm}% bold math5
\def\d{\mathrm{d}}

\def\Re{\text{Re}}
\begin{document}
\title{Image theory for a sphere with negative permittivity}

\date{\today}
\author{Matt R. A. Maji\'c}
\email{mattmajic@gmail.com}
\affiliation{The MacDiarmid Institute for Advanced Materials and Nanotechnology,
School of Chemical and Physical Sciences, \\
Victoria University of Wellington,
PO Box 600, Wellington 6140, New Zealand}

\begin{abstract}
An image system for a point charge outside a dielectric sphere is presented for all complex values of relative permittivity $\epsilon=\epsilon'+i\epsilon''$. The standard image integral solution of a point charge outside a dielectric sphere involving an image point charge plus a line source is shown to diverge for $\epsilon'<-1$, and a correction is proposed for this case, involving image multipoles of infinite magnitude that regularise the divergent line integral. The number of these multipoles depends on the position of $\epsilon$ relative to the resonant values $\epsilon=-1-1/n$ for positive integer $n$. The internal potential and dipole sources are also considered. 
\end{abstract}

\maketitle

\section{Introduction}
The image system for a point charge near a dielectric sphere has been solved by Heaviside and various authors in the late 20th century \cite{poladian1988,lindell1992,norris1995charge}, as a point charge at the inversion point plus some charge distribution on the line segment from the origin to the inversion point. The image theory has been used for example to analyse the interactions of two dielectric spheres \cite{poladian1988,lindell1993twospheres}. However, the image integral diverges for relative permittivity $\epsilon=\epsilon'+i\epsilon''$ with $\epsilon'<-1$, even if the total potential is finite, and this problem has not been addressed. While in electrostatics negative permittivity does not exist physically (although this may be possible to achieve with static charge configurations \cite{yan2013negative}), metals such as gold and silver show negative permittivity at optical frequencies, and for nano-particles a quasi-static analysis is applicable. Quasi-static here means that the spatial dependence of the electric field may be approximated with zero wavenumber so that the wave equation reduces to Laplace's equation, but the permittivity corresponding to a certain non-zero frequency is used.
%While the line integral is not always ideal for computation, this integral has been approximated as a number of point charges at different locations, where the number is increased for a desired accuracy \cite{cai2007fmm}. \color{blue} Can this approach be extended to a regularised integral?\color{black}

In nano-photonics, quasi-static solutions are used to derive simple resonance conditions for particles small compared to the wavelength. The problem of a point dipole near a sphere with negative permittivity applies to surface enhanced Raman spectroscopy involving dipolar molecules and metallic nano-particles which show permittivity with large negative real part at optical frequencies. For a sphere, there is a resonance for each spherical multipole order $n=1,2,3...$, and the quasi-static resonances occur at $\epsilon=\epsilon_\infty=-1-1/n$. These are also known as plasmon resonances, where the electric field intensity near the surface of the sphere becomes very large, and this is exploited to detect single molecules \cite{LeRu2009}. In particular, for silver in water, the relative permittivity at low optical frequencies has $\epsilon'<-1$ and $\epsilon''\lesssim0.2$. 
%However, for $\epsilon'<-1$ but $\epsilon\neq\epsilon_\infty$ the potential is finite but the image/integral form of the potential diverges. 

The document is organised as follows. In section \ref{corrected} 
we regularize the image/integral solution so that it reproduces the series solution for $\epsilon'<-1$, by adding a finite number of infinite-magnitude multipoles. These multipoles are confirmed visually by plots computed using the spheroidal harmonic solution \cite{majic2017super} which converges in all space and can be used to plot the analytic continuation of the external potential inside the sphere. The internal potential is then treated, and in section \ref{dipoles} results are summarised for dipole sources.

\section{Corrected/regularized image for negative permittivity} \label{corrected}

\begin{figure}[!htb]
(a)\includegraphics[scale=.45]{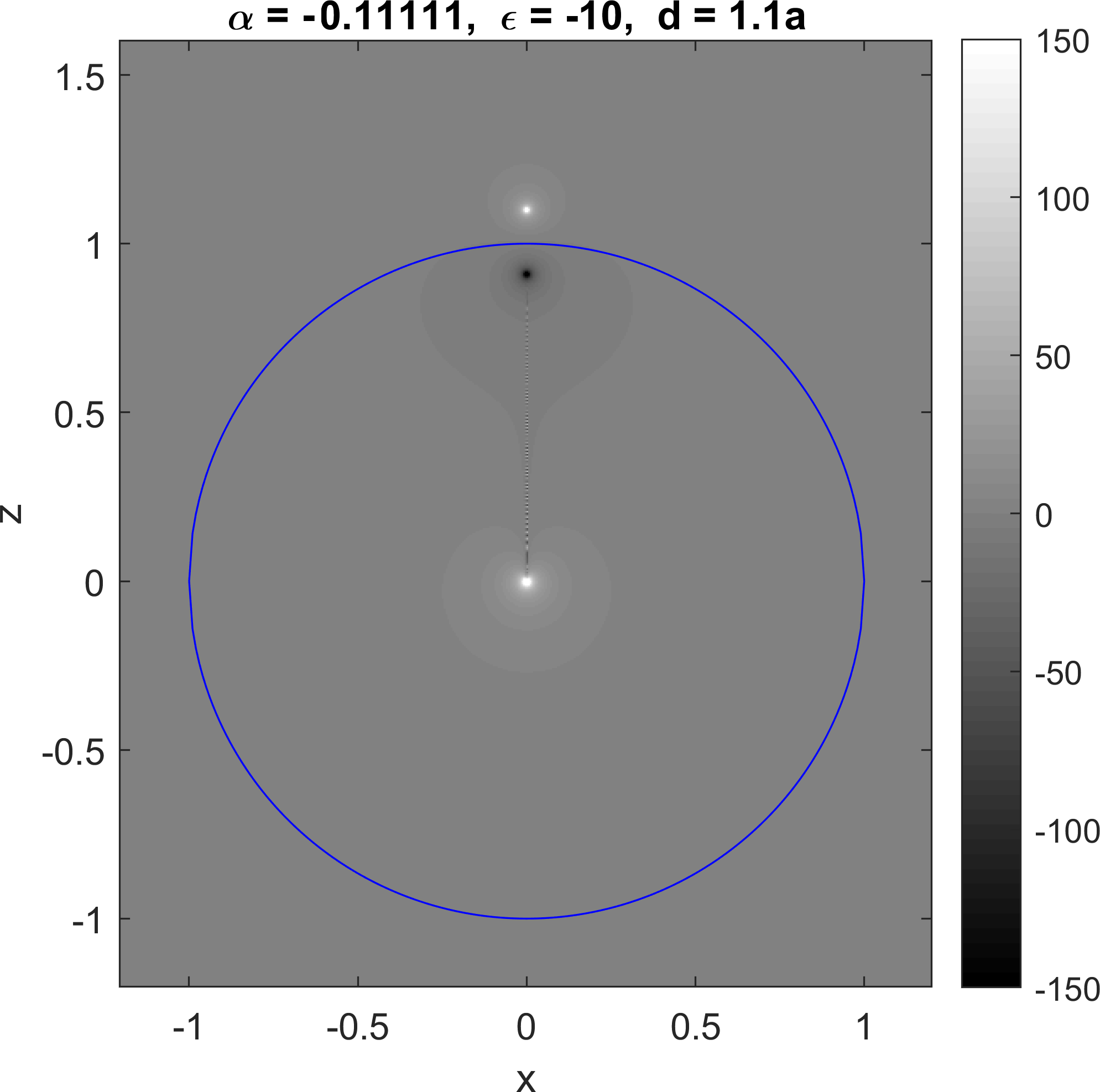} 
(b)\includegraphics[scale=.45]{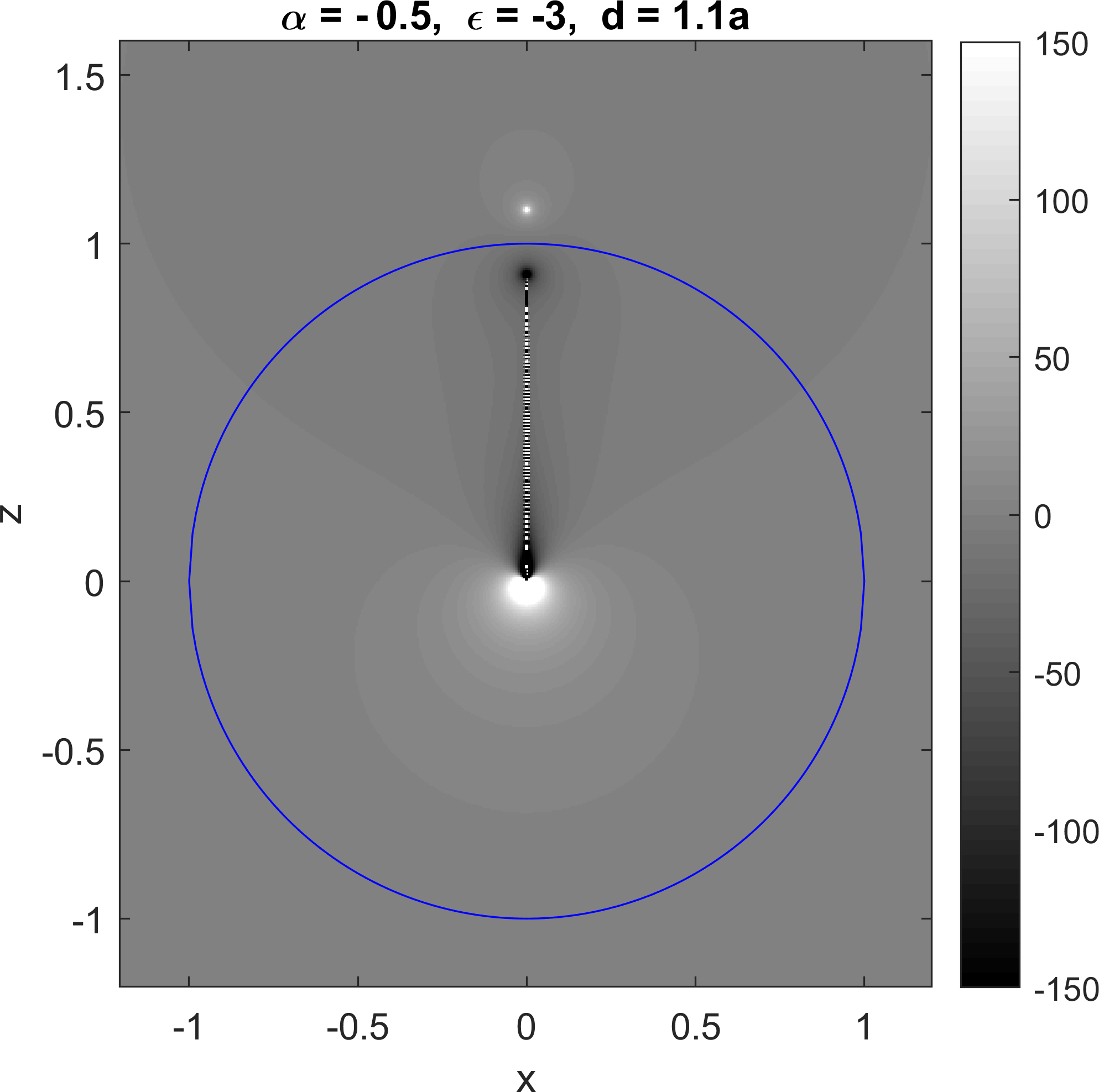}
(c)\includegraphics[scale=.45]{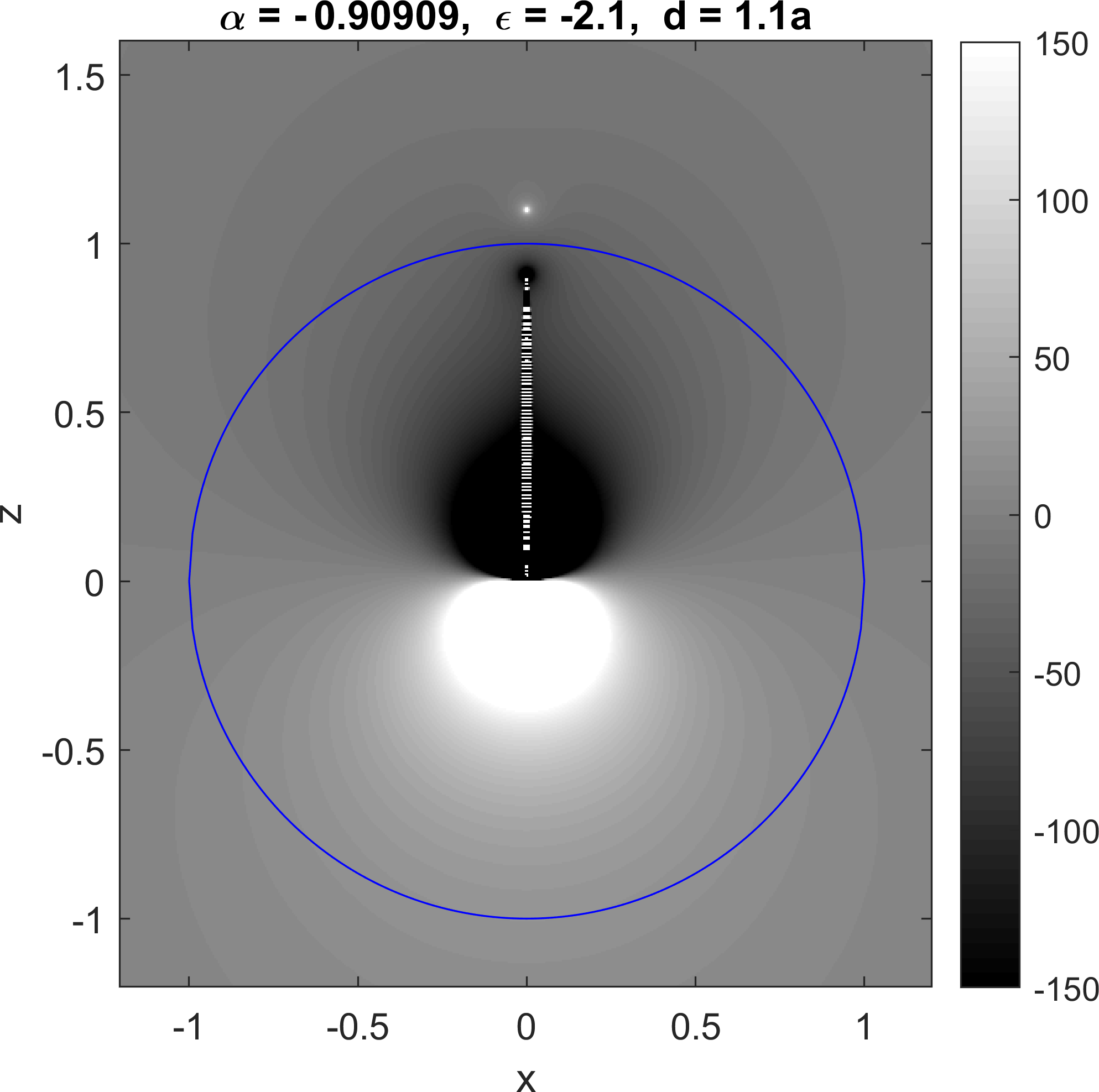}\\
(d)\includegraphics[scale=.45]{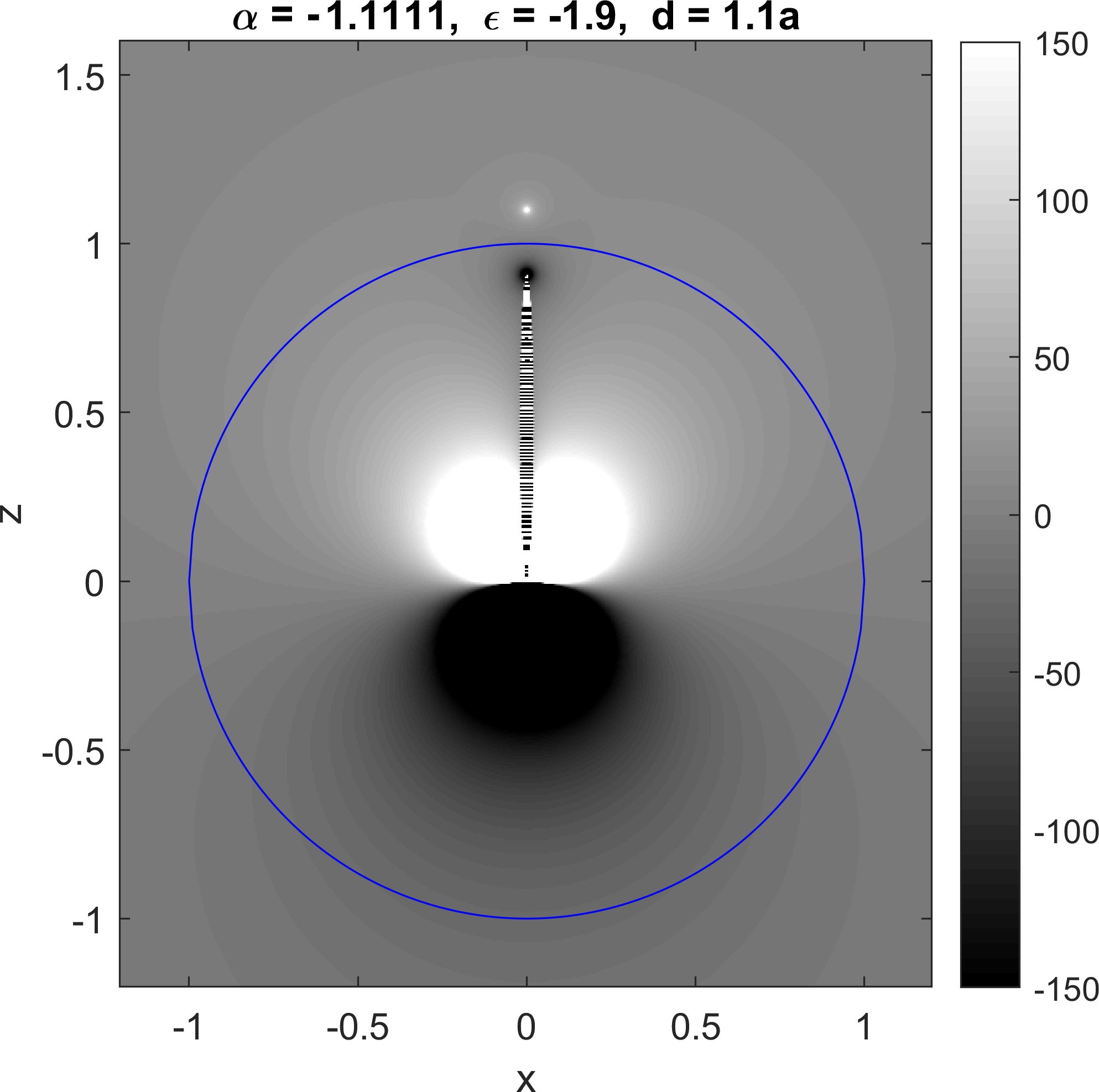}
(e)\includegraphics[scale=.45]{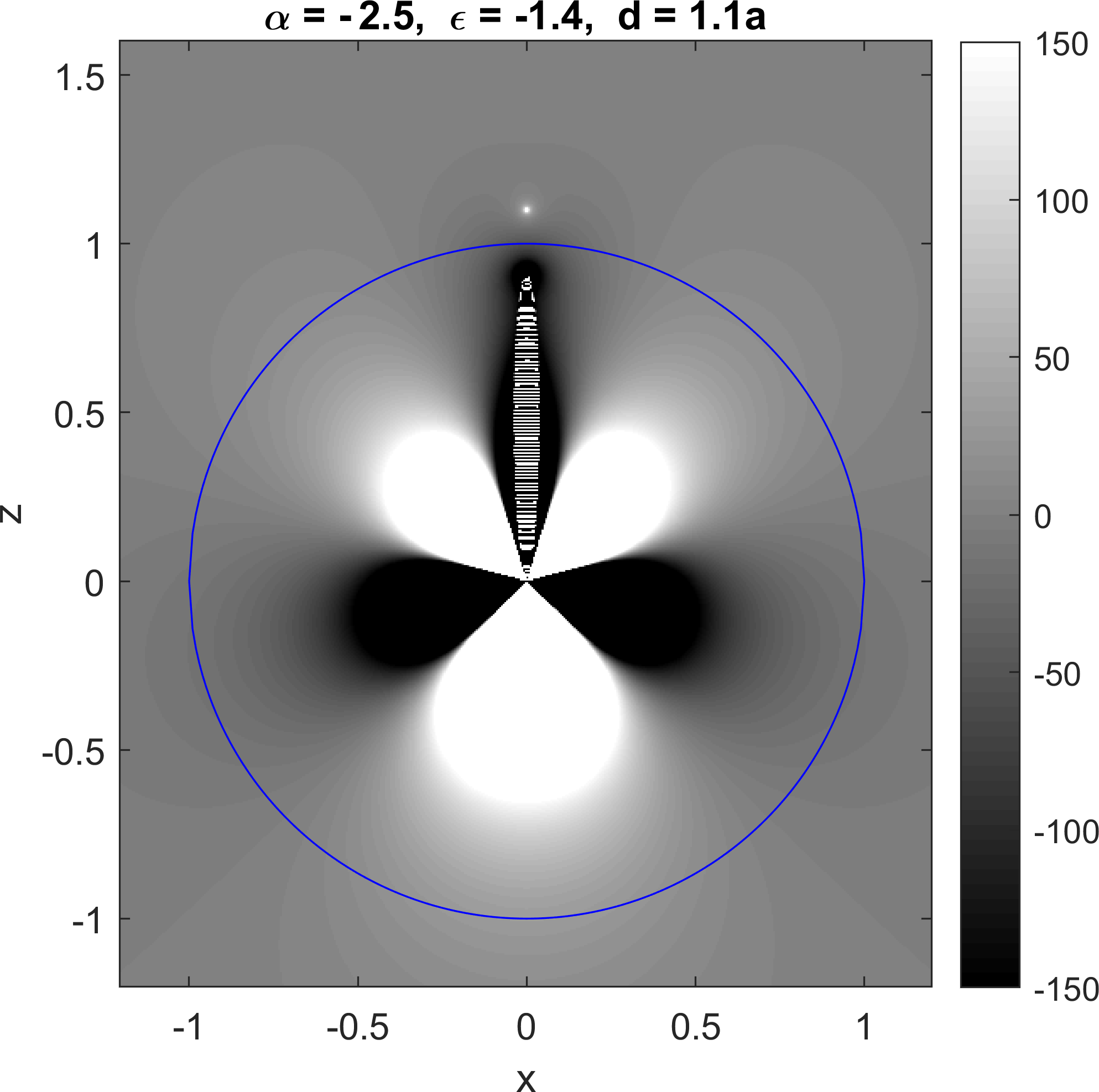}
(f)\includegraphics[scale=.45]{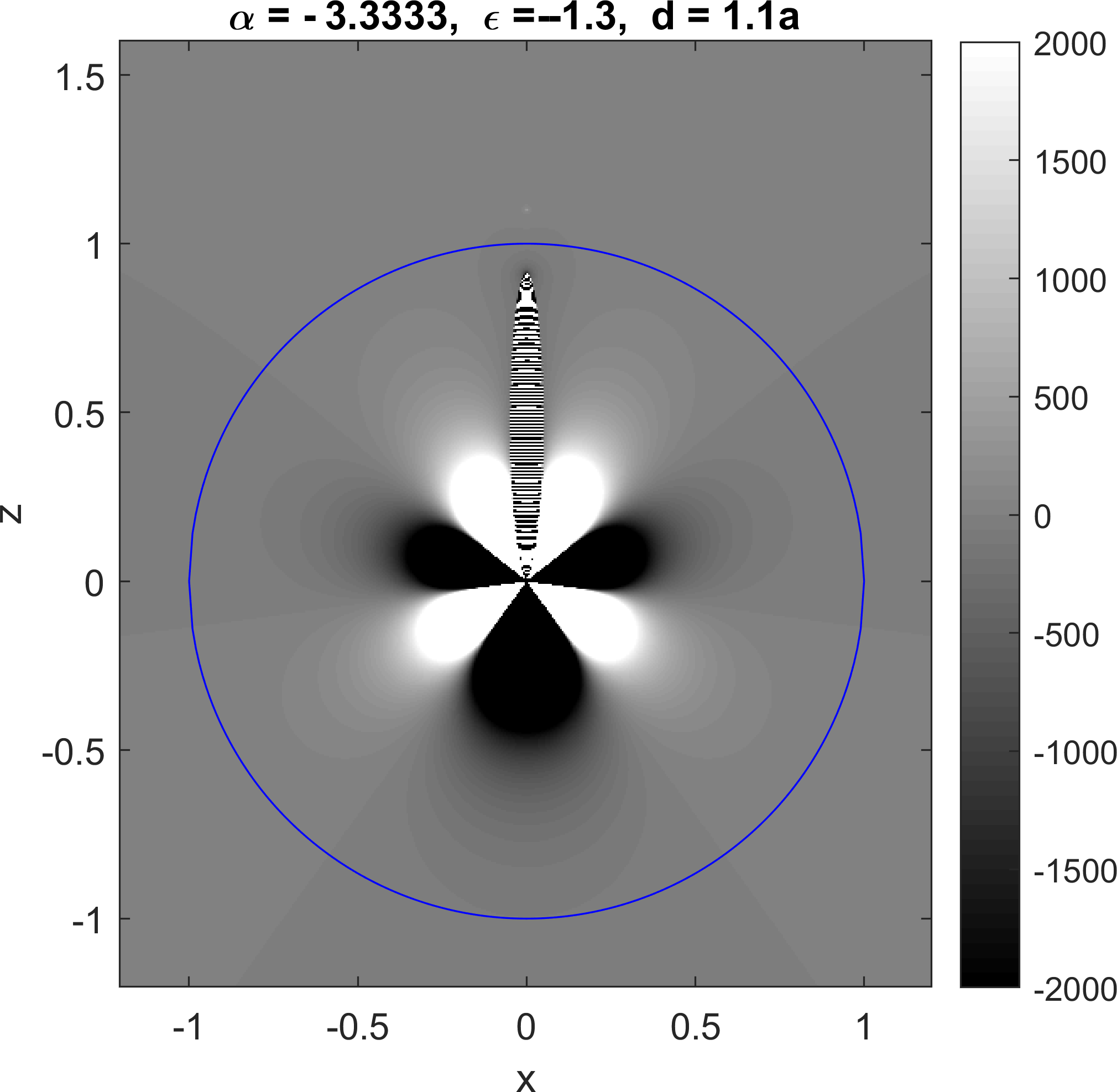}
\caption{Plots of the analytic continuation of the potential $V_s+V_e$ for various $\epsilon$, as computed by the spheroidal harmonic series \ref{VsQP} with a truncation order of $n=300$. The greyscale has been truncated for clarity. The series converges very slowly near the image line for $\epsilon\rightarrow-1$, creating the thin striped spheroidal region which should be ignored.  For $\epsilon=1.9$ (d), there are actually 2 positive lobes either side of the image line. } \label{plots}
\end{figure}

Consider a point charge $Q$ located at $z=d$ above a sphere radius $a$, permittivity $\epsilon_s$ in a medium with permittivity $\epsilon_m$. The relative permittivity is $\epsilon=\epsilon_s/\epsilon_m$. Using spherical $(r,\theta,\phi)$ and cylindrical $(\rho,z,\phi)$ coordinates, the exciting potential is 
\begin{align}
V_e=V_0\frac{a}{r_d}; \qquad 
r_d=\sqrt{\rho^2+(z-d)^2}, \qquad
V_0=\frac{Q}{4\pi\epsilon_m a}.
\end{align}
For $\epsilon'>-1$, the solution for the scattered potential $V_s$ is given in \cite{lindell1992} in both series and integral/image form. We separate the image point charge at $z=b=a^2/d$ from the series solution:
\begin{align}
V_s&=  V_0 \frac{\epsilon-1}{\epsilon+1}\bigg[-\frac{b}{r_b} +  \sum_{n=0}^\infty \frac{1}{n(\epsilon+1)+1}\bigg(\frac{b}{r}\bigg)^{n+1} P_n(\cos\theta)\bigg] \\
&= V_0 \frac{\epsilon-1}{\epsilon+1}\bigg[-\frac{b}{r_b} + \frac{b}{\epsilon+1} \int_0^1 \frac{u^{\alpha-1}\d u}{\sqrt{\rho^2+(z-bu)^2}} \bigg]\\
\text{where }r_b&=\sqrt{\rho^2+(z-b)^2} , \qquad \alpha = \alpha'+i\alpha'' = \frac{1}{\epsilon+1}. \label{Vs1}
\end{align}
The equality of the series and integral expressions comes from recognising the generating function for the Legendre polynomials in the integrand:
\begin{align}
\frac{1}{\sqrt{\rho^2+(z-bu)^2}}=\sum_{n=0}^\infty \frac{b^nu^n}{r^{n+1}}P_n(\cos\theta),
\end{align}
so that the integrand is a simple series of the powers $u^{n+\alpha-1}$.
However, if $\alpha'<0$ ($\epsilon'<-1$) this integrand will contain terms which diverge for $n<-\alpha'$, due to the antiderivative being infinite at the endpoint $u=0$. In fact the series and integral forms would be equal if we ignored the bottom limit, but we cannot do this if we want to make physical sense of the integral as a charge distribution. To counter this problem, we subtract off this infinite part. For $\alpha'<0,~\alpha\neq0,-1,-2,-3...$, (case $\alpha=0$ is finite but should be treated separately) we may write
\begin{align}
\sum_{n=0}^\infty \bigg(\frac{b}{r}\bigg)^{n+1}\frac{1}{n+\alpha} P_n(\cos\theta)=\lim_{\nu\rightarrow0} \left[b\int_\nu^1 \frac{u^{\alpha-1}\d u}{\sqrt{\rho^2+(z-bu)^2}} + \sum_{n=0}^{\lfloor-\alpha\rfloor} \frac{\nu^{n+\alpha}}{n+\alpha}\bigg(\frac{b}{r}\bigg)^{n+1} P_n(\cos\theta)\right]. 
\end{align}
Where $\lfloor x\rfloor$ is the floor of $x$, defined here as the largest integer $\leq \Re\{x\}$. As $\nu\rightarrow0$, the integral and series both diverge, but their difference is finite and equal to the integral evaluated at the top end point which is equal to the left hand side. This approach of regularisation has been used recently for the Stieltjes transformation, and discussed in more detail in \cite{galapon2017problem} (see their eq 2.18).
The image consists of an image point charge at the inversion point, plus a line charge, plus, if $\alpha'<0$ ($\epsilon'<-1$), a finite number ($\lfloor-\alpha+1\rfloor$) of infinite-magnitude multipoles at the origin that regularise the divergence of the line source. For $\alpha=-1/2$ for instance (top centre of figure \ref{plots}), an image point charge at the origin is introduced with opposite sign to the line charge distribution (same sign as the source charge). 
Then the correct form of the potential for $\alpha\neq0,-1,-2...$ is 
\begin{align}
V_s
&= V_0 \frac{\epsilon-1}{\epsilon+1}\Bigg[-\frac{b}{r_b} + \frac{1}{\epsilon+1} \lim_{\nu\rightarrow0}\bigg(b\int_\nu^1 \frac{u^{\alpha-1}\d u}{\sqrt{\rho^2+(z-bu)^2}} +\sum_{n=0}^{\lfloor-\alpha\rfloor} \frac{\nu^{n+\alpha}}{n+\alpha}\bigg(\frac{b}{r}\bigg)^{n+1} P_n(\cos\theta) \bigg)\Bigg] \label{Vs corrected}
\end{align}
When $\alpha'>0$ ($\epsilon'>-1$), the sum vanishes and the expression reduces to \eqref{Vs1}. The concept of a divergent charge distribution is very common in electrostatics - for example it is used to describe an ideal dipole, which is the limit of two opposite charges placed infinitely close together while their charge magnitude diverges so that the electric field is finite.\\

\begin{figure}[!htb]
(a)\includegraphics[scale=.455]{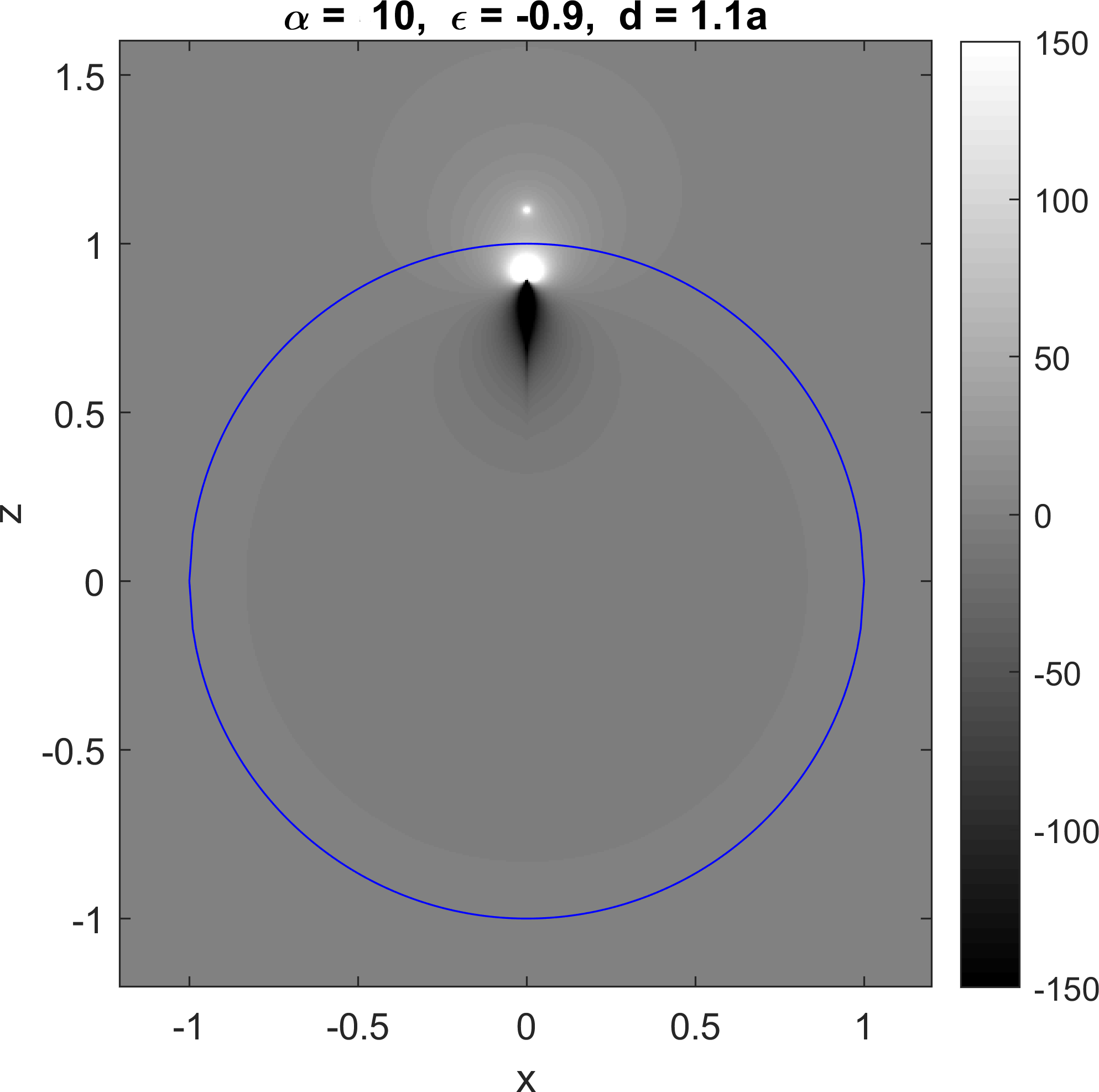} 
(b)\includegraphics[scale=.455]{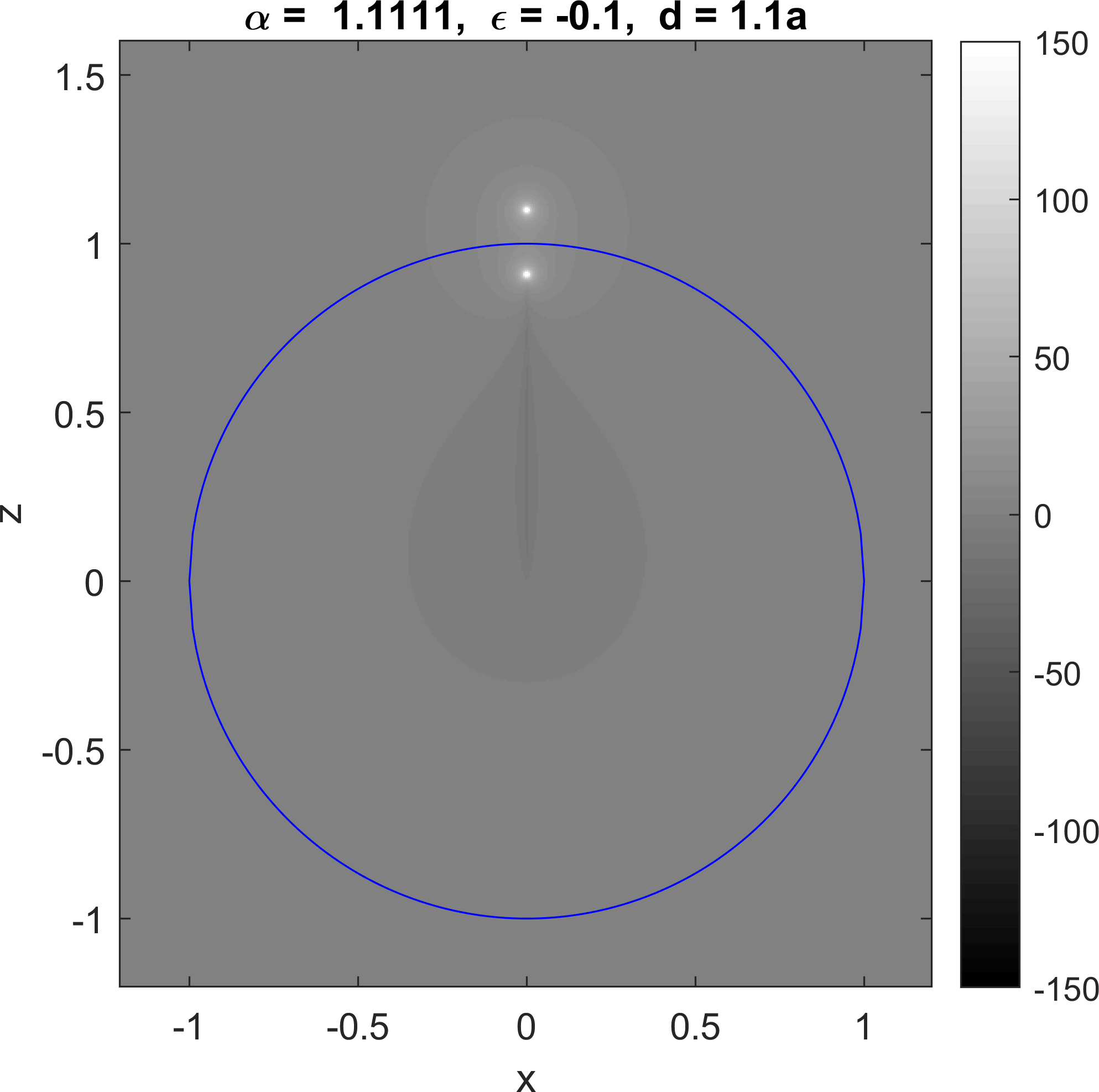}
(c)\includegraphics[scale=.455]{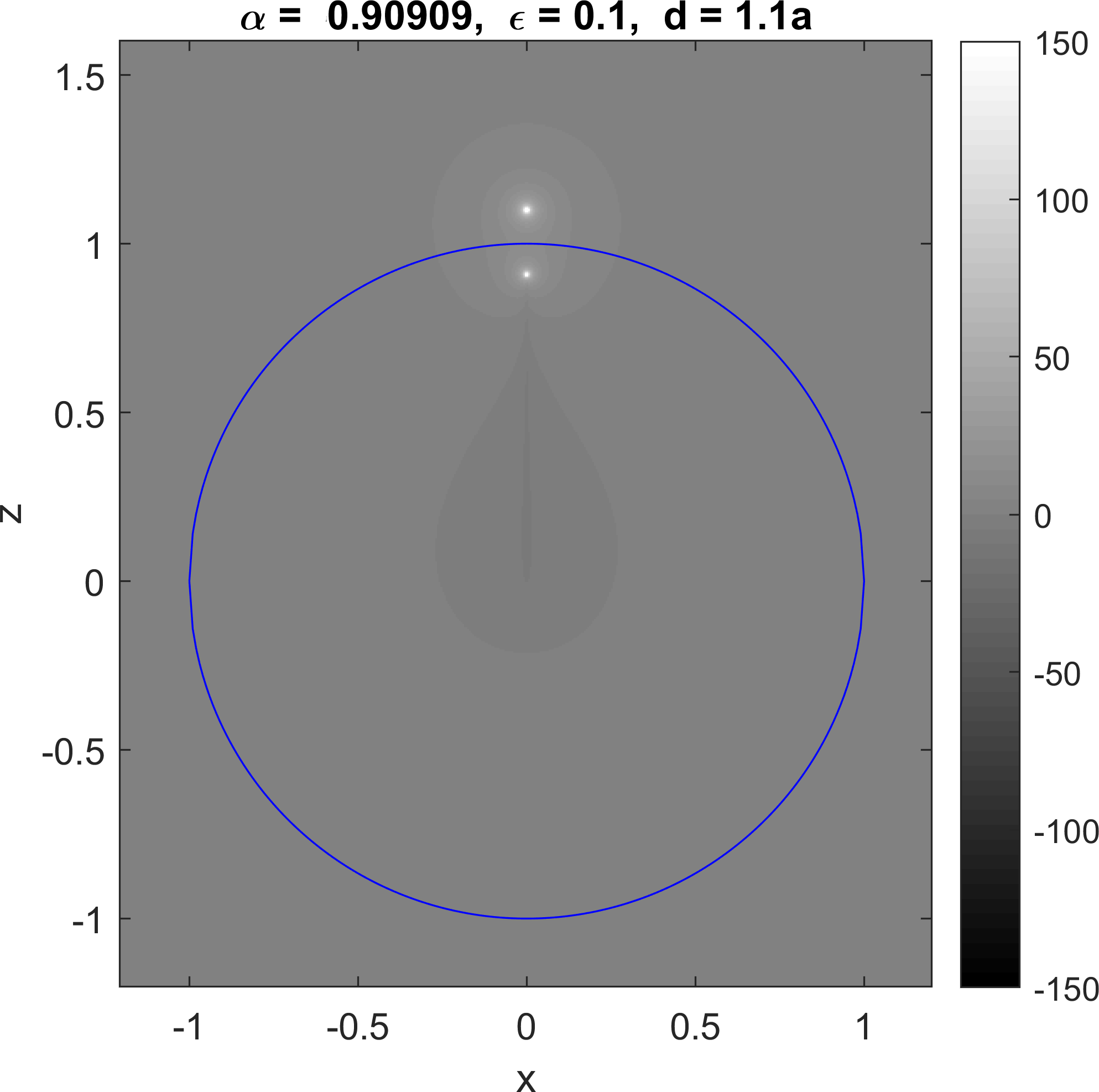}\\
(d)\includegraphics[scale=.455]{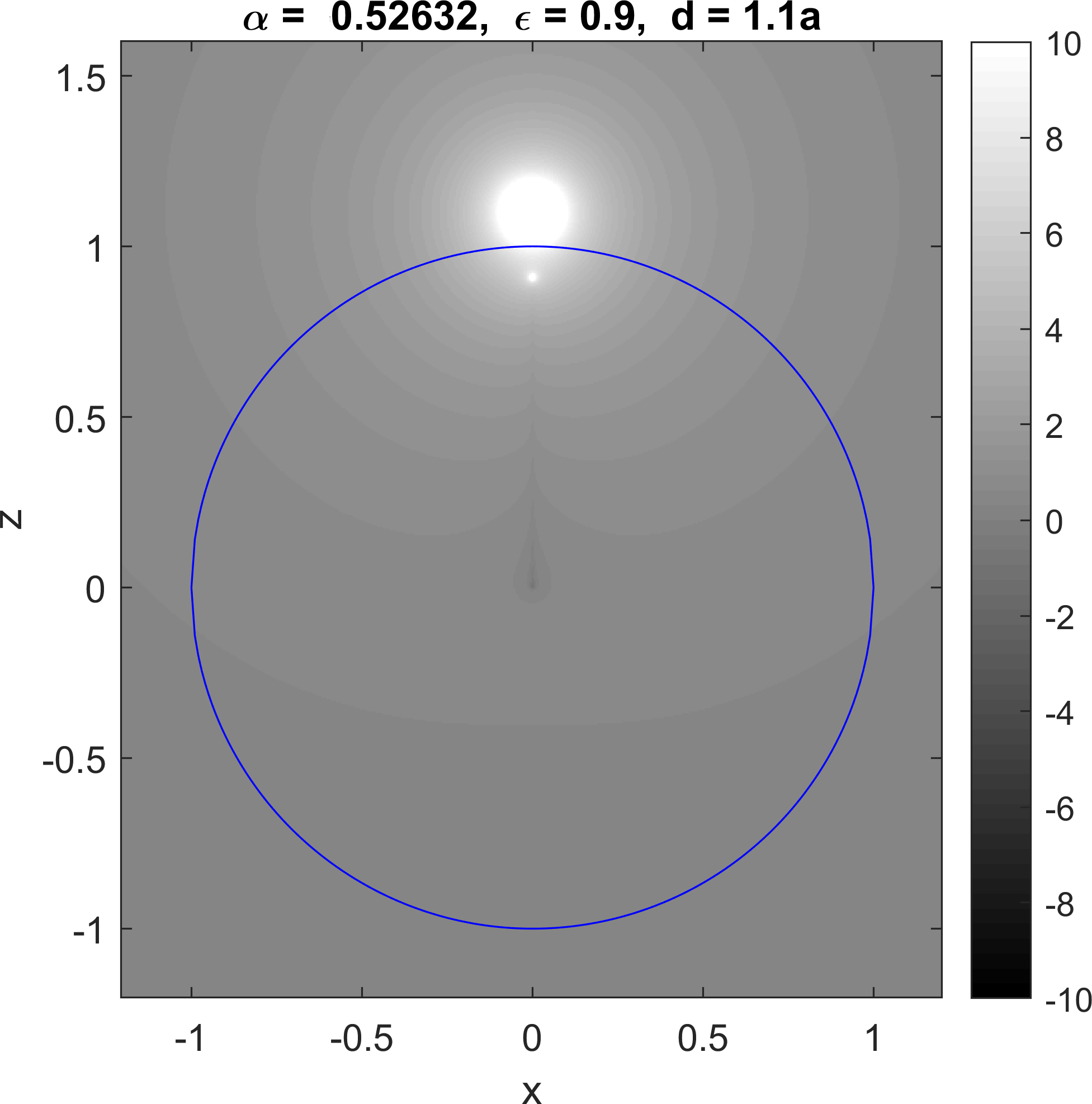}
(e)\includegraphics[scale=.455]{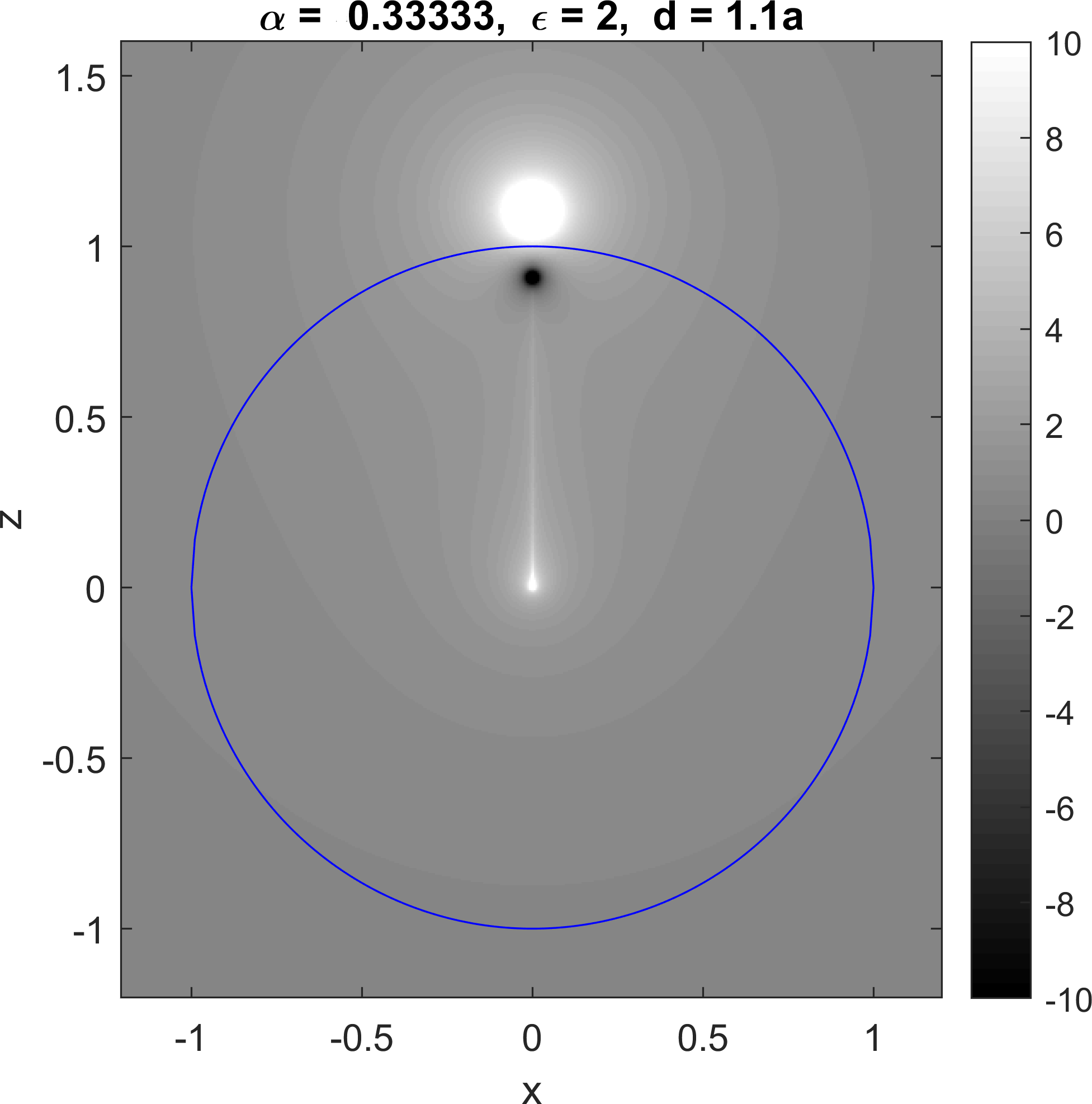}
(f)\includegraphics[scale=.455]{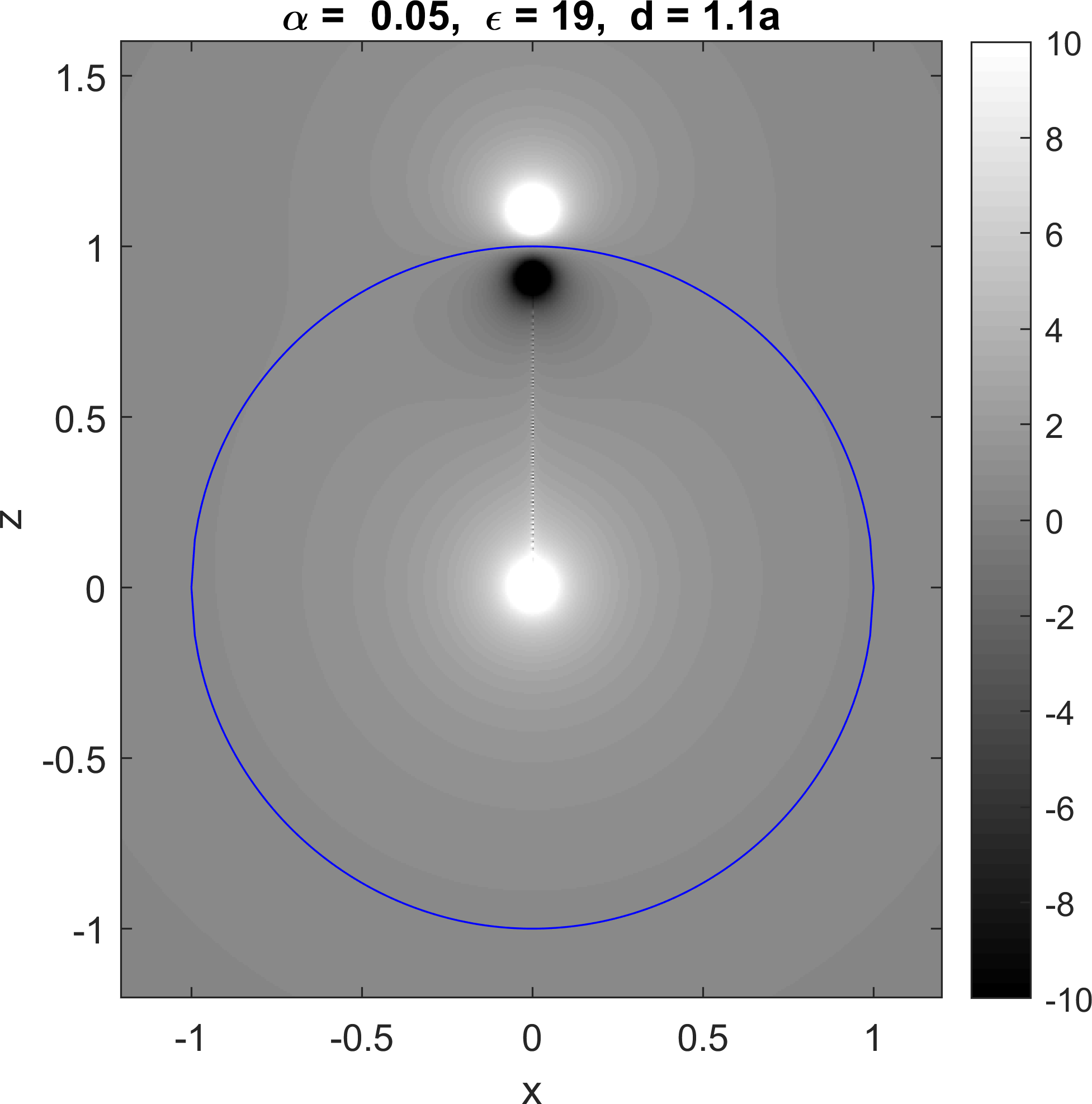}
\caption{ Plots of the analytic continuation of the potential $V_s+V_e$ for various $\epsilon>-1$. Here there are no infinite multipoles. Note the lower intensity scale in (d)-(f).} \label{plots reg}
\end{figure}

An interesting case is $\epsilon=-1$ ($|\alpha|=\infty$), where the limit as $\epsilon\rightarrow-1$ from below oscillates rapidly, but from above the image can be obtained from the spherical harmonic series solution:
\begin{align}
V_s(\epsilon=-1)
&= 2V_0 \sum_{n=0}^\infty n \bigg(\frac{b}{r}\bigg)^{n+1}P_n(\cos\theta) = 2V_0 \frac{b^2}{r_b^2}\cos\theta_b 
\end{align}
where $\theta_b$ is the colatitude from the inversion point. This is a dipole at the inversion point pointing towards the source (seen the top left of figure \ref{plots reg} for $\epsilon=-0.9$).
And for the conducting sphere, $|\epsilon|=\infty$ in any direction in the complex plane ($\alpha=0$):
\begin{align}
V_s(|\epsilon|=\infty)
&= -V_0 \sum_{n=0}^\infty \bigg(\frac{b}{r}\bigg)^{n+1}P_n(\cos\theta) = -V_0 \frac{b}{r_b}.
\end{align}
%which is the standard image charge solution for the conducting sphere. Note that this solution is the same for any phase of $\epsilon$ in the complex plane. 
The form of the image \eqref{Vs corrected} can be verified by plotting the spheroidal harmonic series solution \cite{majic2017super} which converges everywhere except the image line, for all $\alpha\neq-1,-2,-3,...~$:
\begin{align}
V_s&=V_0 \frac{\epsilon-1}{\epsilon+1}\bigg[-\frac{b}{r_b} + 2 \sum_{n=0}^\infty (2n+1)\prod_{k=1}^n\frac{\alpha-k}{\alpha+k}~Q_n(\bar\xi)P_n(\bar\eta) \bigg] \label{VsQP} \\
& \text{where } \bar{\xi}=\frac{r+r_b}{b}, \quad \bar\eta=\frac{r-r_b}{b}.\nonumber
\end{align}
The familiar potentials for $\epsilon'>-1$ are plotted in figure \ref{plots reg}, and the potentials for various $\epsilon'<-1$ are plotted in figure \ref{plots}. In both figures $V_s$ is computed with \eqref{VsQP}. The cases $\epsilon\rightarrow\infty$ (bottom right of figure \ref{plots reg}) and $\epsilon\rightarrow-\infty$ (top left of figure \ref{plots}) both produce the identical result of the Kelvin image charge found for the conducting sphere. Figure \ref{plots reg} shows some interesting features even for $\epsilon'>-1$, for example the image system flip sign as $\epsilon$ passes through 1 (see plots for $\epsilon=0.9,2$), but there is no real difference between $\epsilon=0.1$ and $\epsilon=-0.1$. In figure \ref{plots}, as $\epsilon\rightarrow-1$ from below we see the addition of a new infinite multipole at the origin each time $\epsilon$ passes a resonance $\epsilon_\infty=-1-1/n$. There is a drastic difference for $\epsilon$ being slightly either side of each resonance, which is very apparent for the plots with $\epsilon=-2.1$ and $\epsilon=-1.9$. The black and white striped regions are due to the finite truncation of the spheroidal harmonic series (higher orders may be summed to minimise this striped region, but this eventually introduces numerical underflow in $Q_n(\bar\xi)$ away from the line segment). The spheroidal series can in fact be proven to converge in all space except the line segment $\bar\xi=1$ by analysis of the limit of each term as $n\rightarrow\infty$:
\begin{align}
\lim_{n\rightarrow\infty}(2n+1)\prod_{k=1}^n\frac{\alpha-k}{\alpha+k}Q_n(\bar\xi)|P_n(\bar\eta)| ~&\propto (-)^n n \frac{\Gamma(-\alpha+n+1)}{\Gamma(\alpha+n+1)} \bar\xi^{-n}\\
&\propto (-)^n n^{-2\alpha}\bar\xi^{-n}
\end{align}
Which converges for $\bar\xi>1$, although slowly for $\alpha\rightarrow-\infty$ ($\epsilon\rightarrow-1$).\\

%As $\epsilon\rightarrow-1$ from above, the image tends towards an image dipole at the inversion point. 
%For $\epsilon$ near the poles $\epsilon_\infty$, we see that $V_s$ may be come extremely large in its physically valid region outside the sphere.

%For comparison with Lindell \cite{Lindell1992}, the potential is presented in terms of a charge distribution, using the dirac delta representation of the solid spherical harmonics: 
%\begin{align}
%V_s&=\frac{1}{4\pi\epsilon_m}\int_0^b\frac{\varrho(z')\d z'}{\sqrt{\rho^2+(z-z')^2}} \\
%\varrho(z)&=\frac{\epsilon-1}{\epsilon+1} \frac{Q}{a} \Bigg[-b\delta(z-b) + \frac{1}{\epsilon+1}\lim_{\nu\rightarrow0}\bigg[ \left(\frac{z}{b}\right)^{\alpha-1} + \sum_{n=0}^{\lfloor-\alpha\rfloor} \frac{\nu^{n+\alpha}}{n+\alpha} b^{n+1} \frac{(-)^n}{n!}\frac{\pd^n}{\pd z^n} \delta(z)\bigg] \Bigg].
%\end{align}

\begin{figure}
\includegraphics[scale=.5]{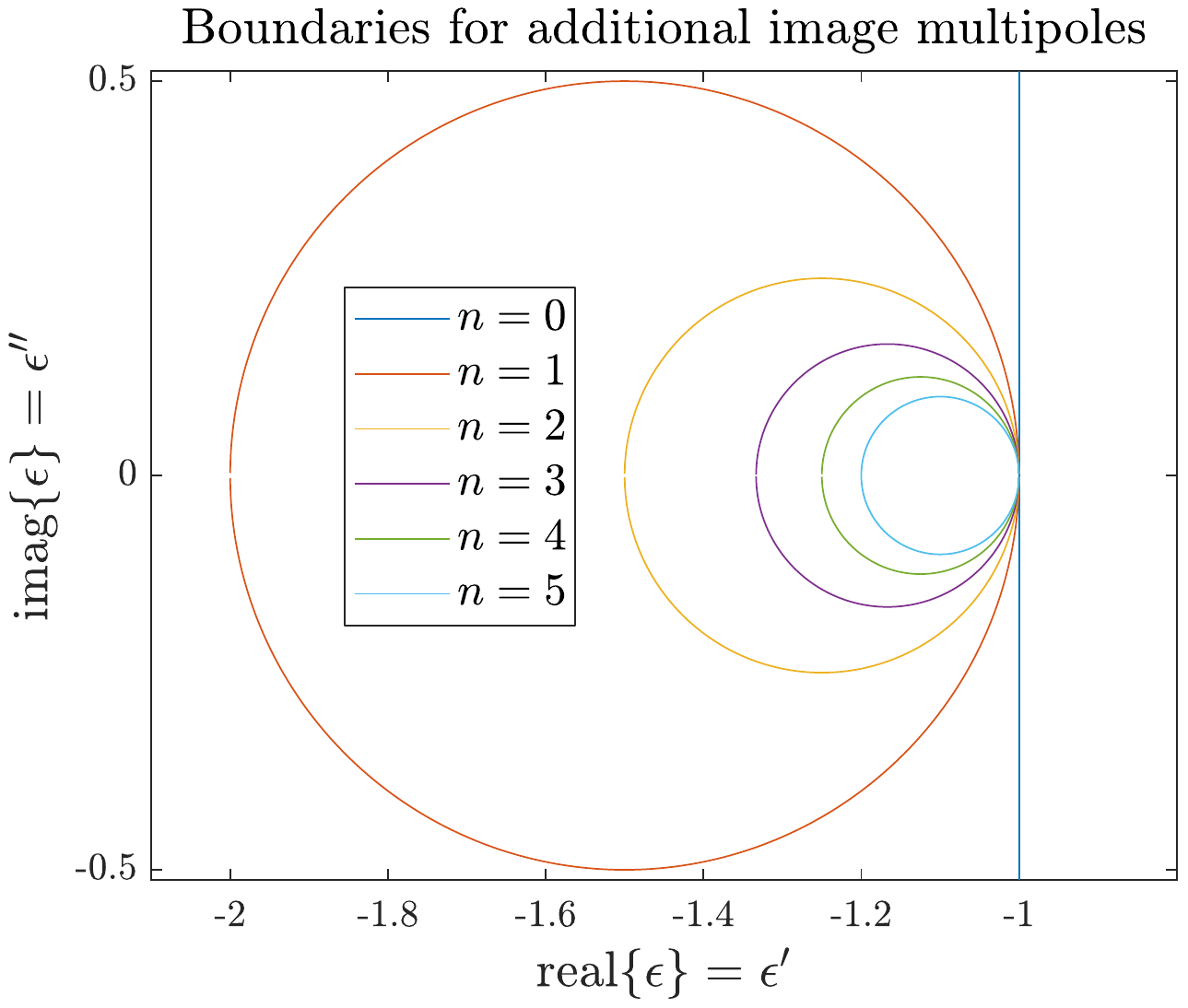}
\caption{Circles that define the number of additional multipoles to be added to the sum in \eqref{Vs corrected}. To the right of $\epsilon'=-1$ there are no additional multipoles, and from there, crossing inside each circle adds one multipole.}\label{epsiloncircles}
\end{figure}

\begin{figure}[!htb]
(a)\includegraphics[scale=.435]{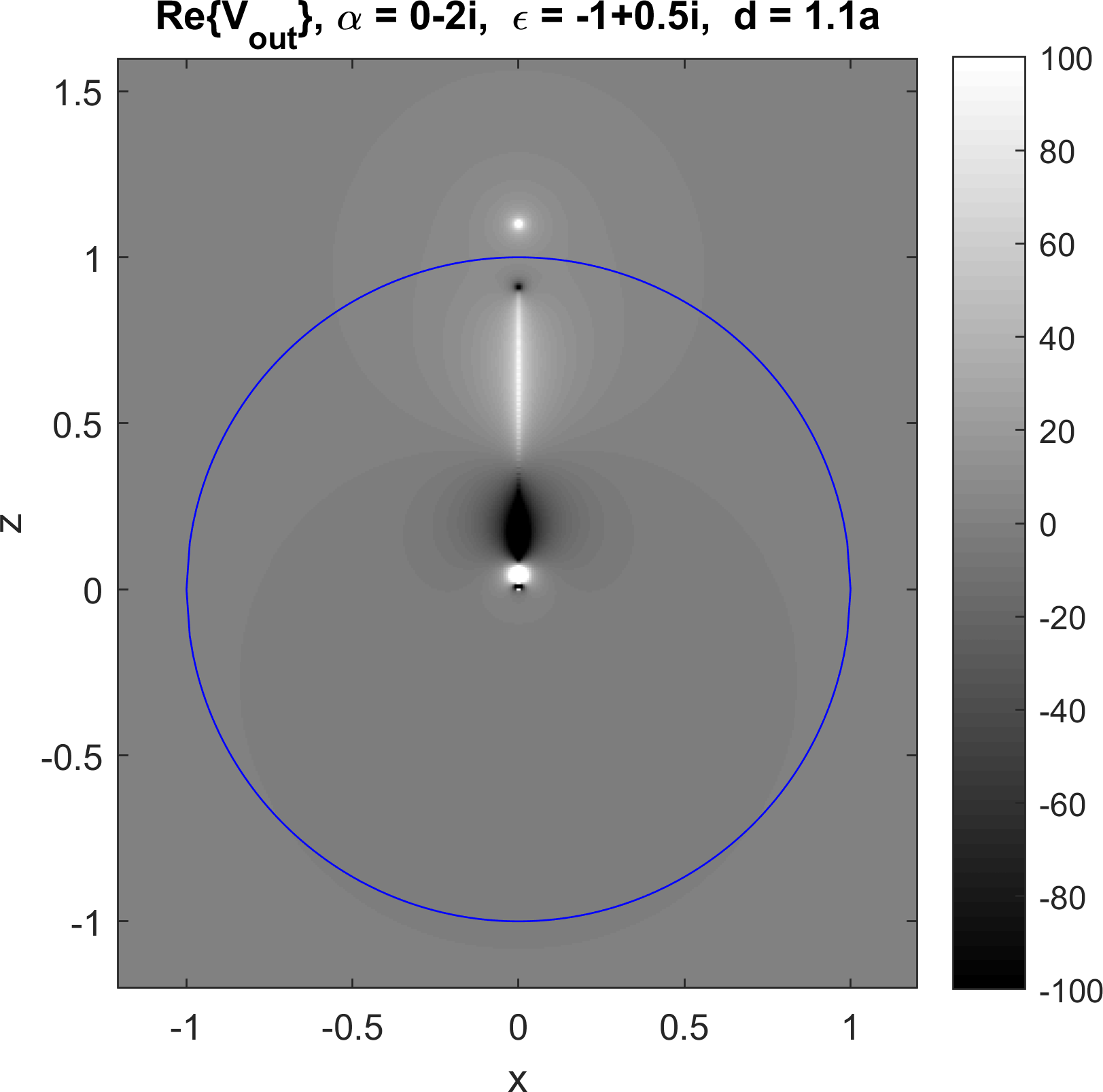} 
(b)\includegraphics[scale=.435]{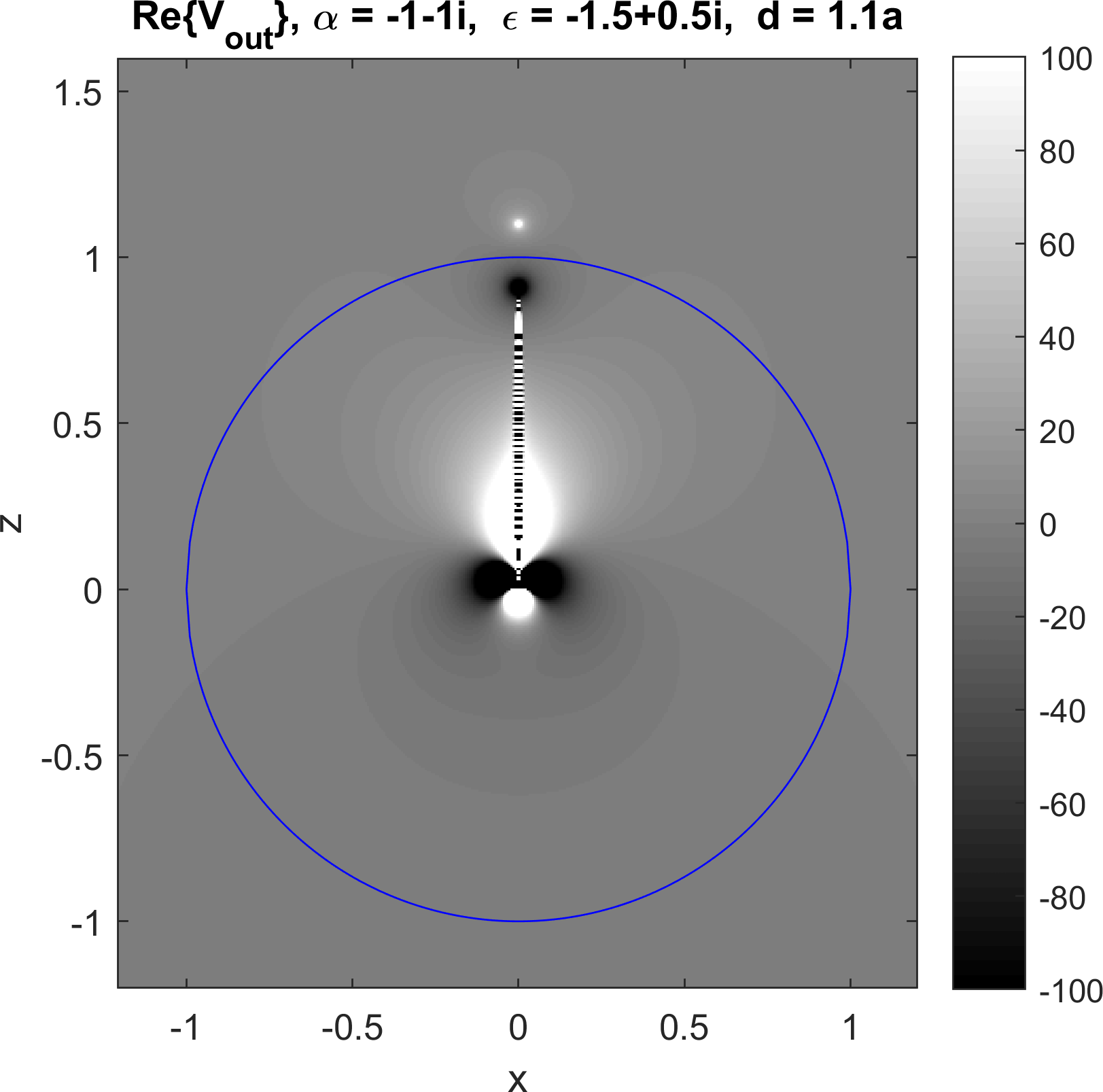} 
(c)\includegraphics[scale=.435]{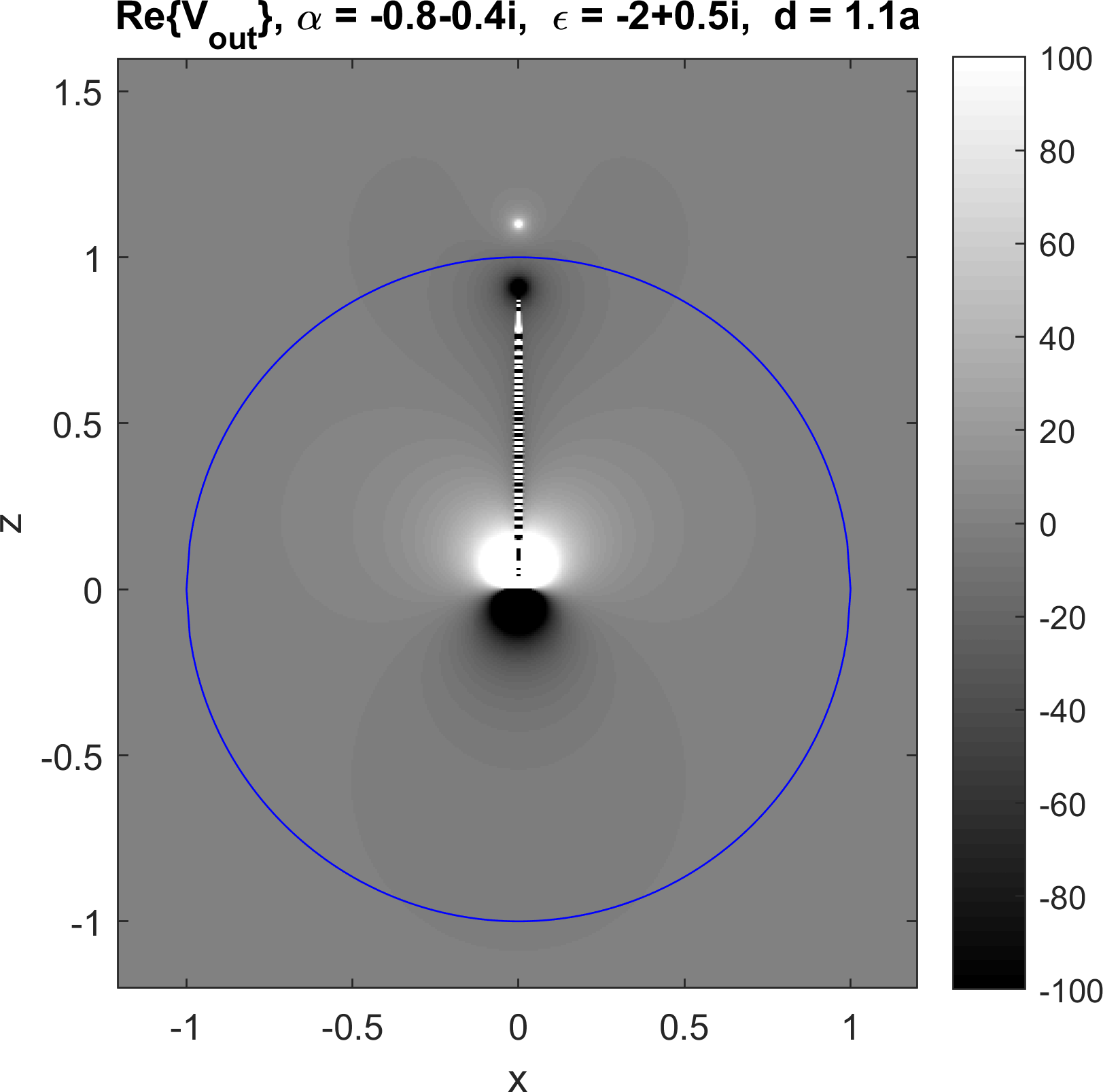} \\
(d)\includegraphics[scale=.435]{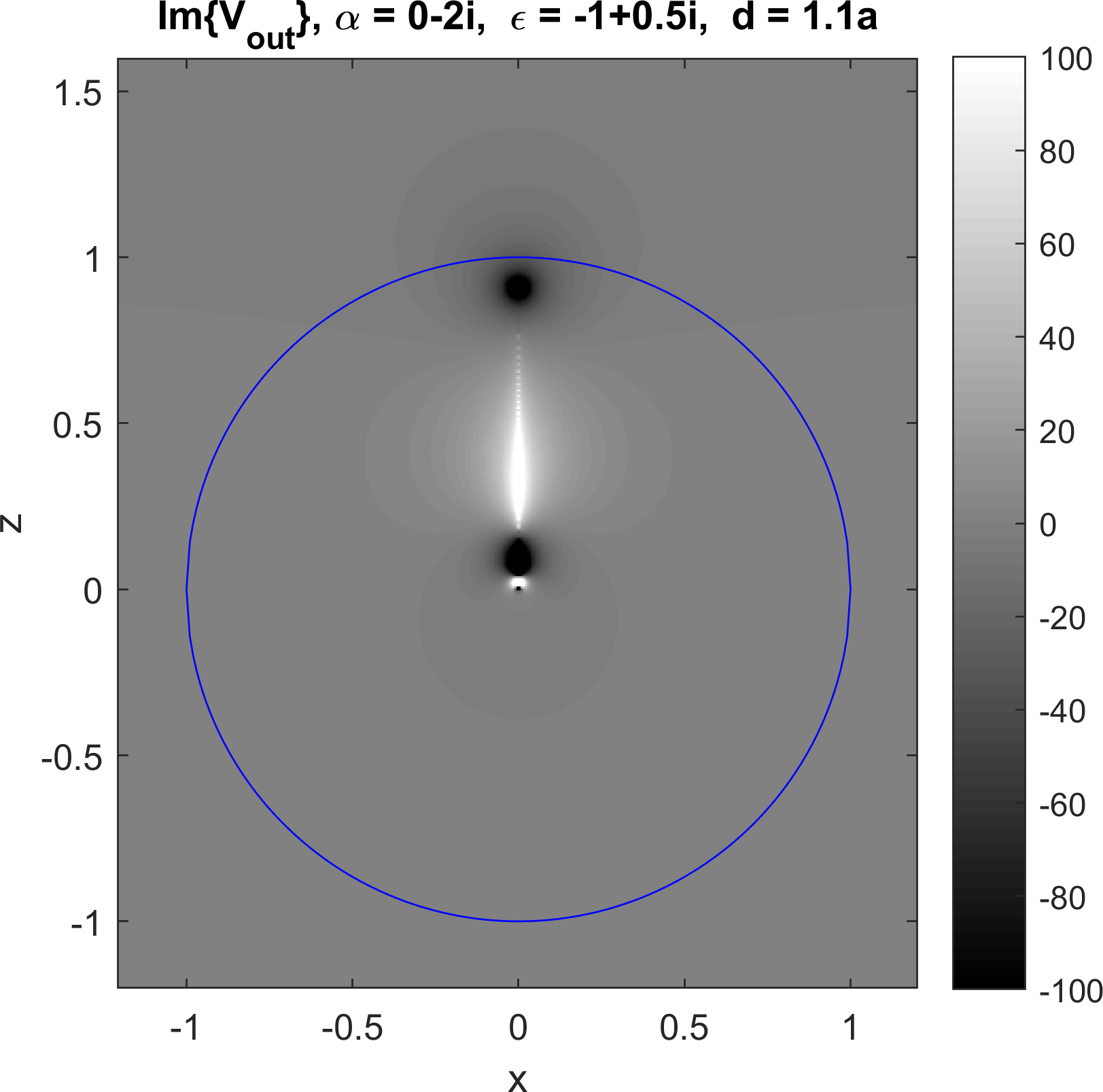} 
(e)\includegraphics[scale=.435]{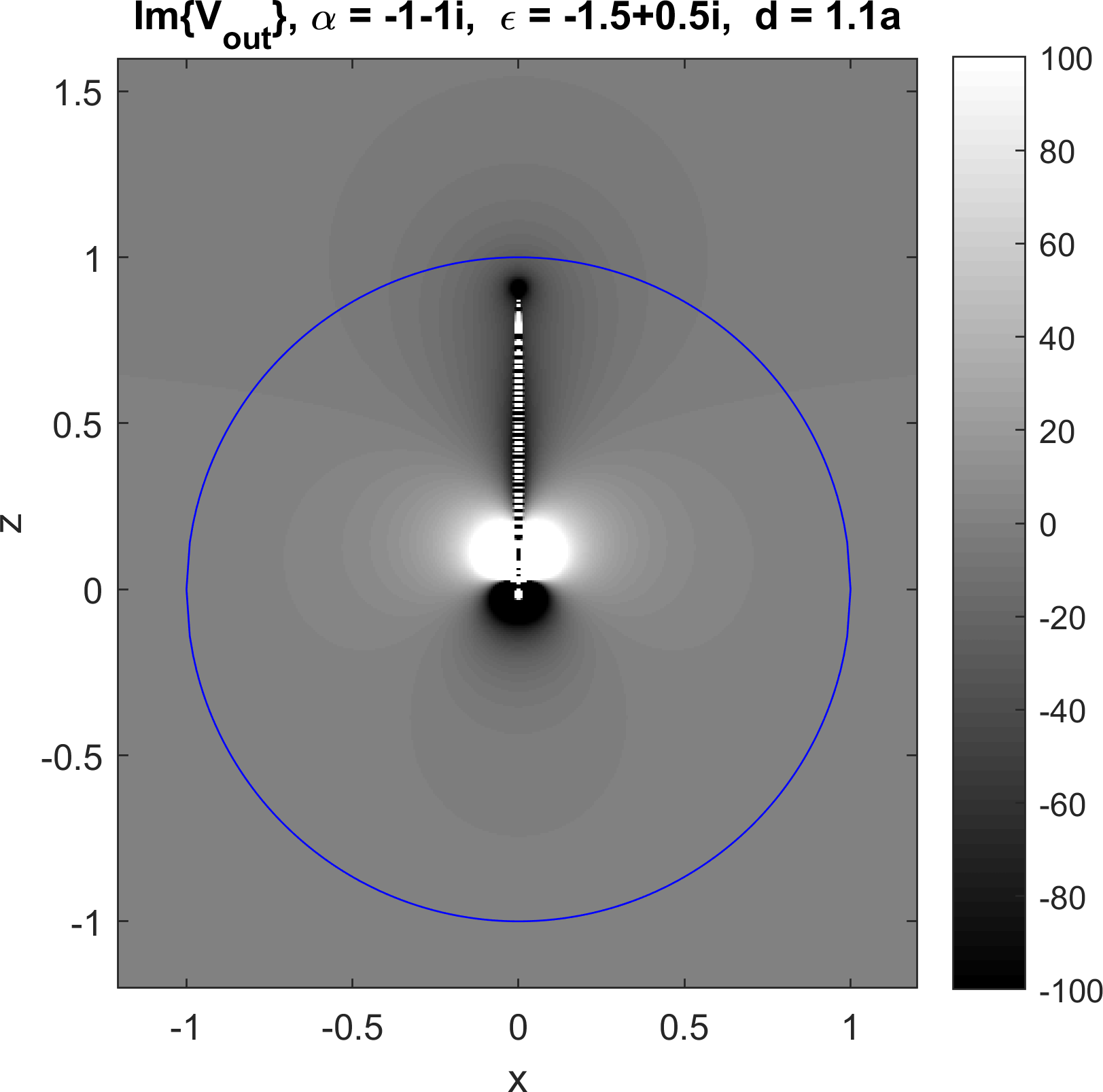} 
(f)\includegraphics[scale=.435]{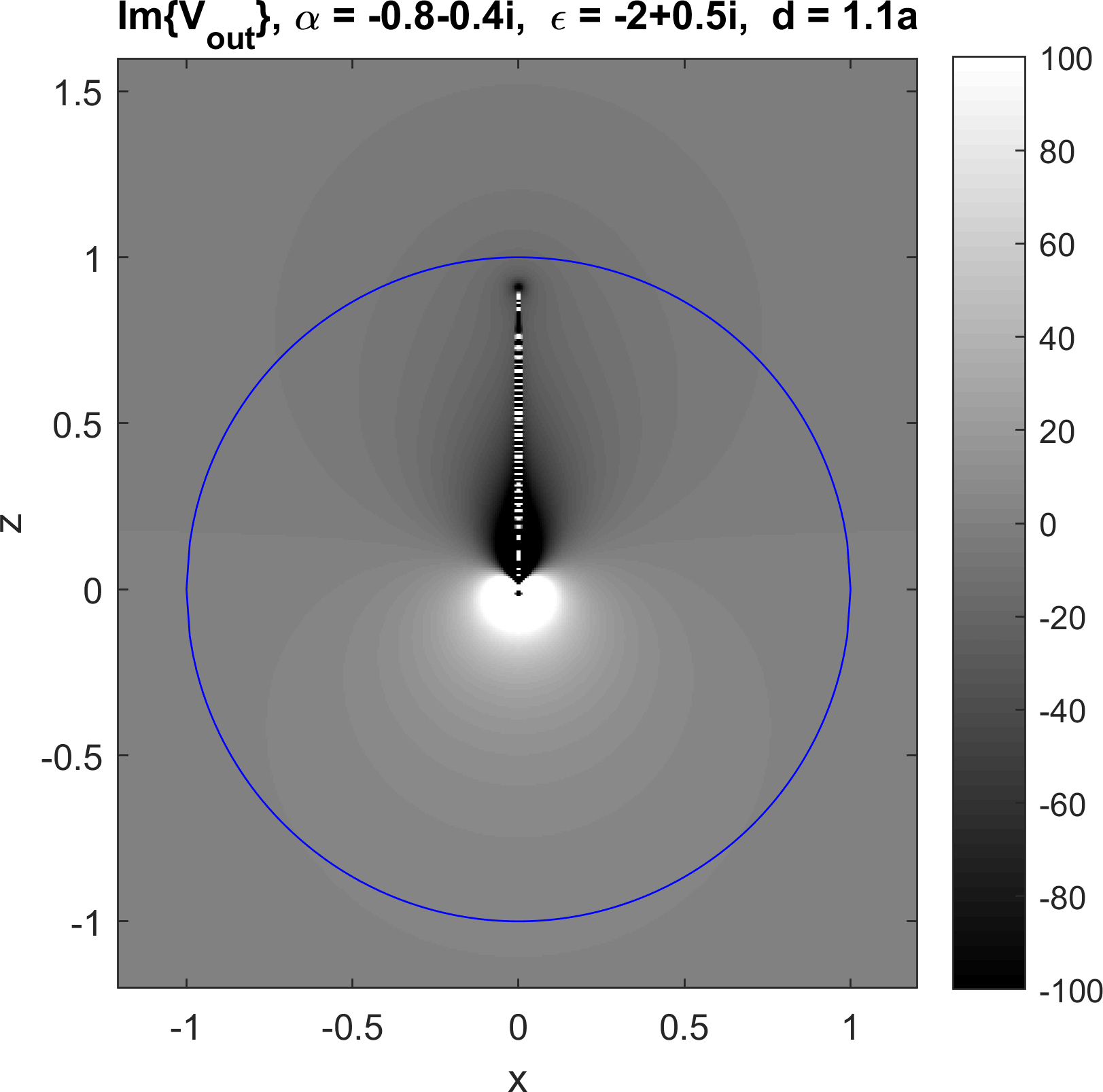} 
\caption{eal and imaginary parts of the analytic continuation of $V_{out}=V_s+V_e$ for complex $\epsilon$.} \label{plots complex}
\end{figure}

For complex $\epsilon$, the real and imaginary parts of $V_s$ are plotted for $\epsilon'=-1,-1.5,-2$, $\epsilon''=0.5$ in figure \ref{plots complex}. As mentioned above, the number of infinite-magnitude multipoles increases by one as $\alpha'$ crosses left of $-n$. The conditions $\alpha'=-n$ are equivalent to $\epsilon''^2=-(\epsilon'+1)/n-(\epsilon'+1)^2$, which are equations for circles in the complex plane for each $n$, shown in figure \ref{epsiloncircles}. For $\epsilon''\neq0$ the potential changes smoothly as $\epsilon$ crosses these boundaries.

\begin{figure}[!htb]
(a)\includegraphics[scale=.451]{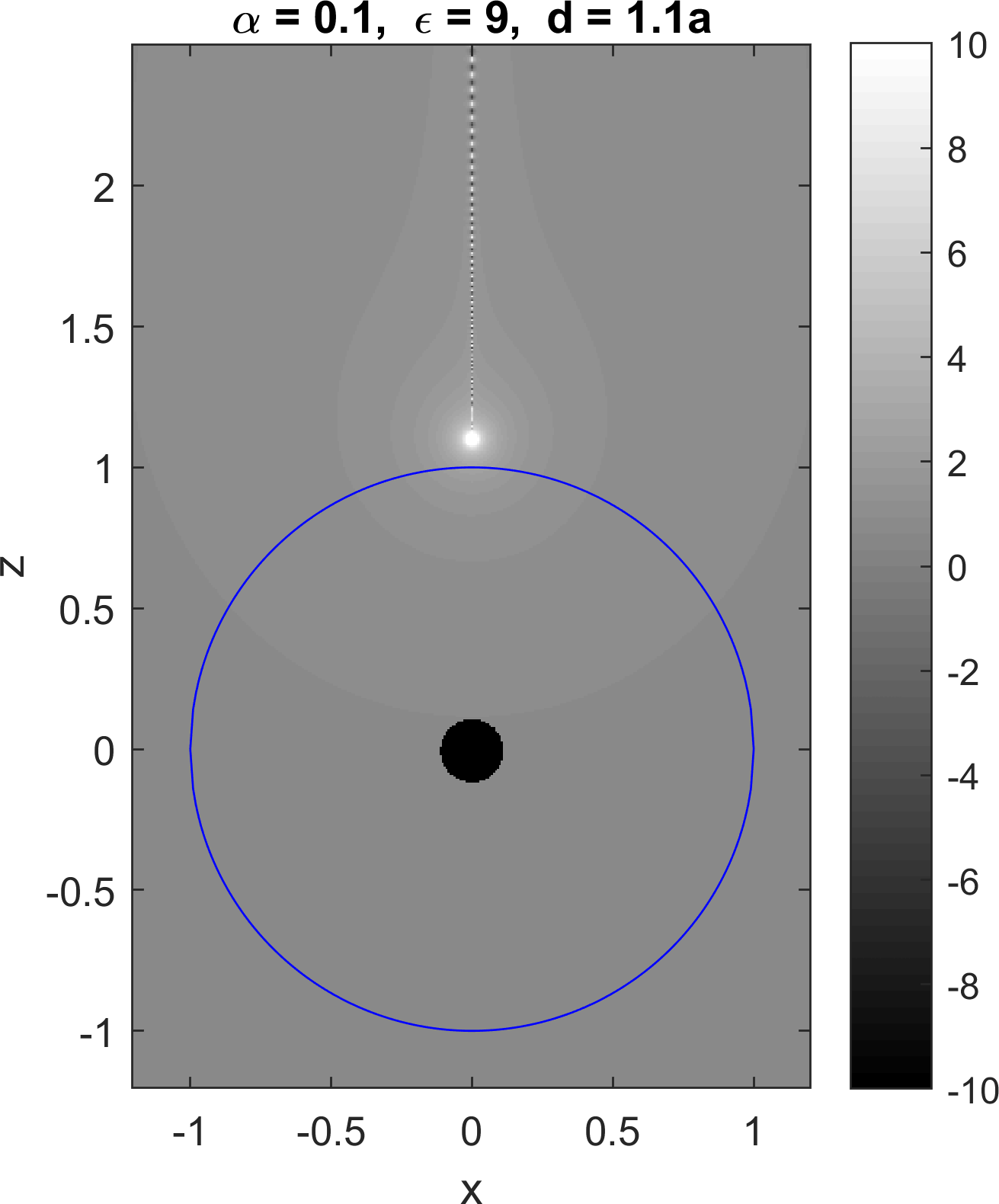}
(b)\includegraphics[scale=.451]{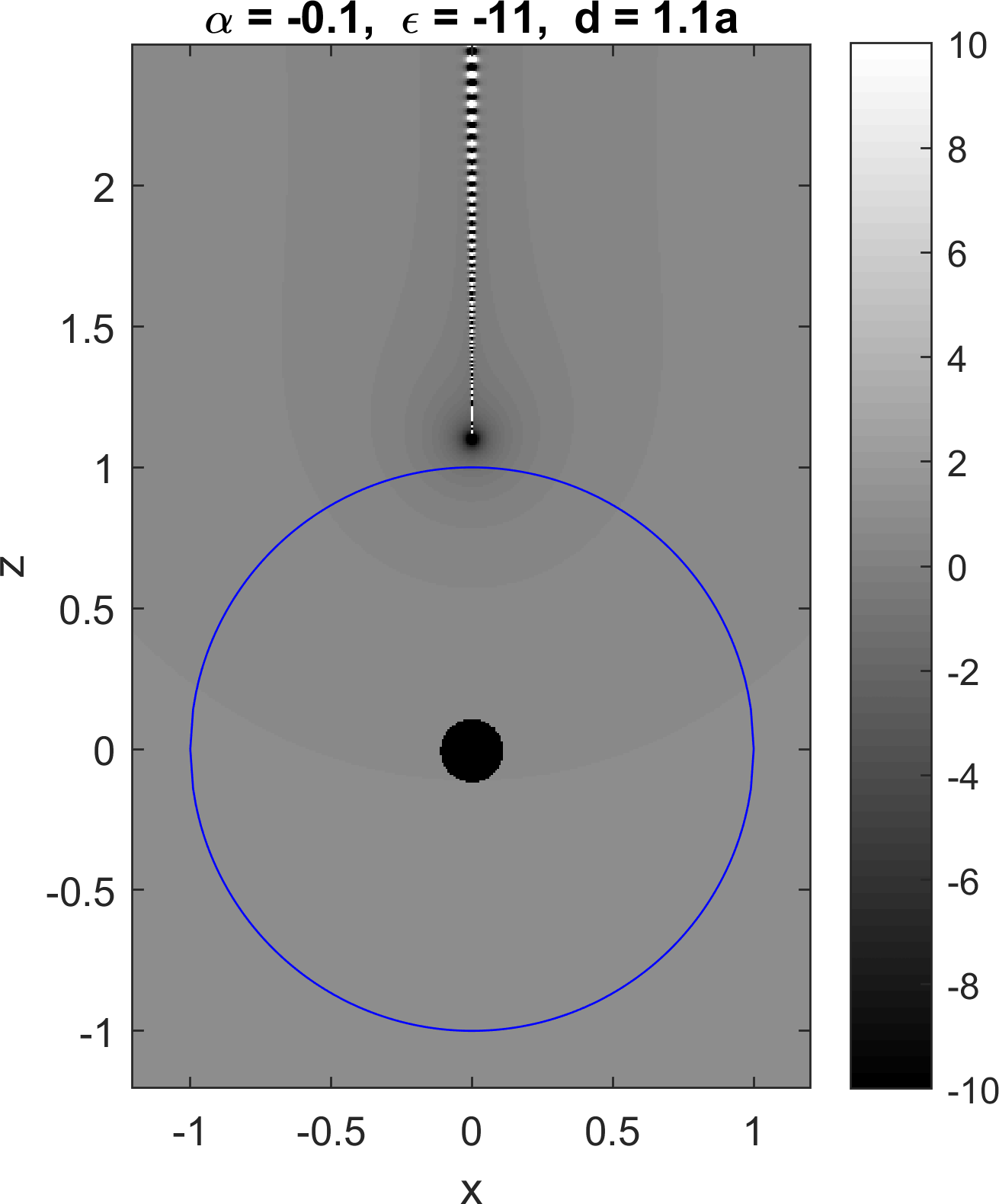}
(c)\includegraphics[scale=.451]{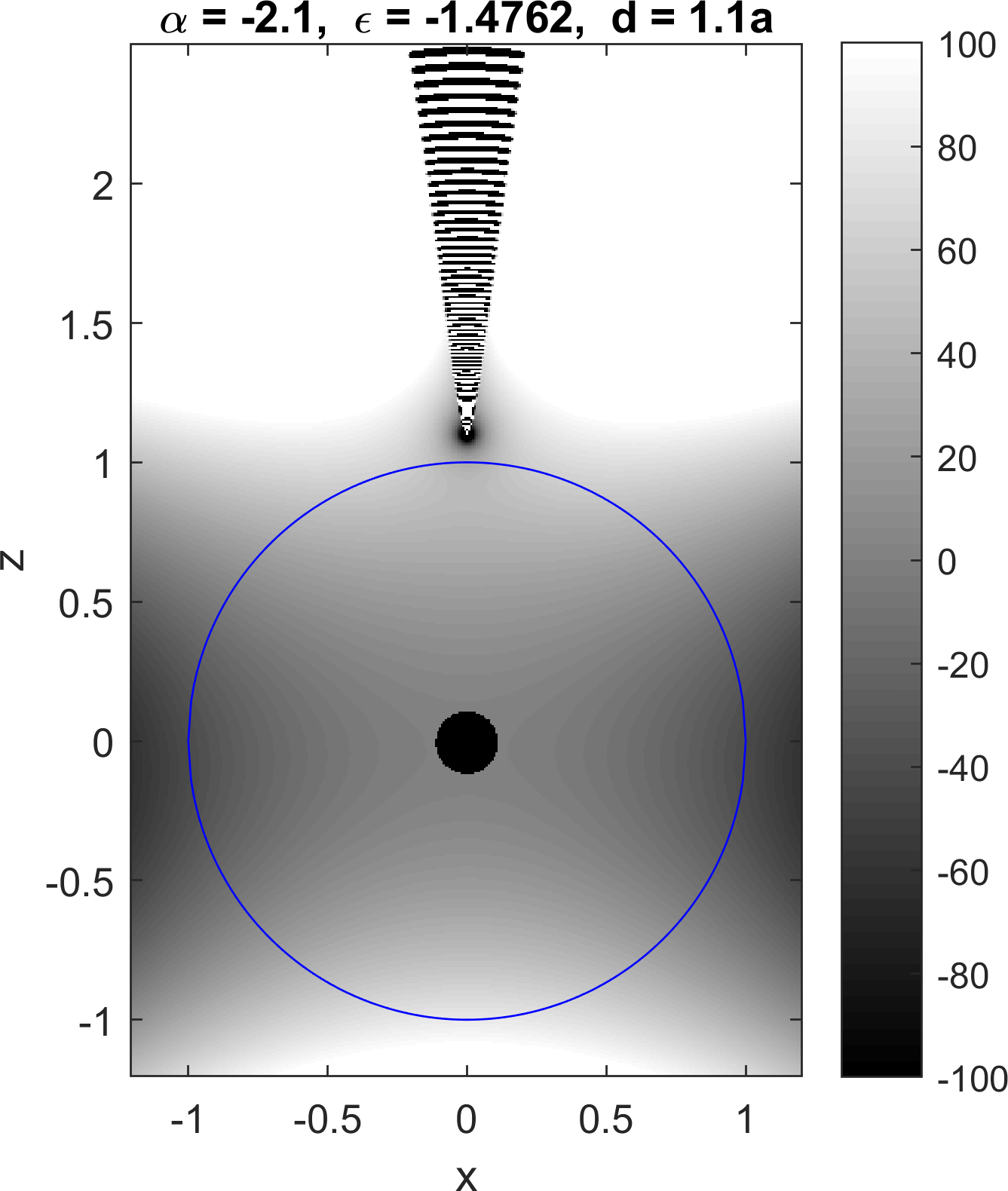}\\
(d)\includegraphics[scale=.451]{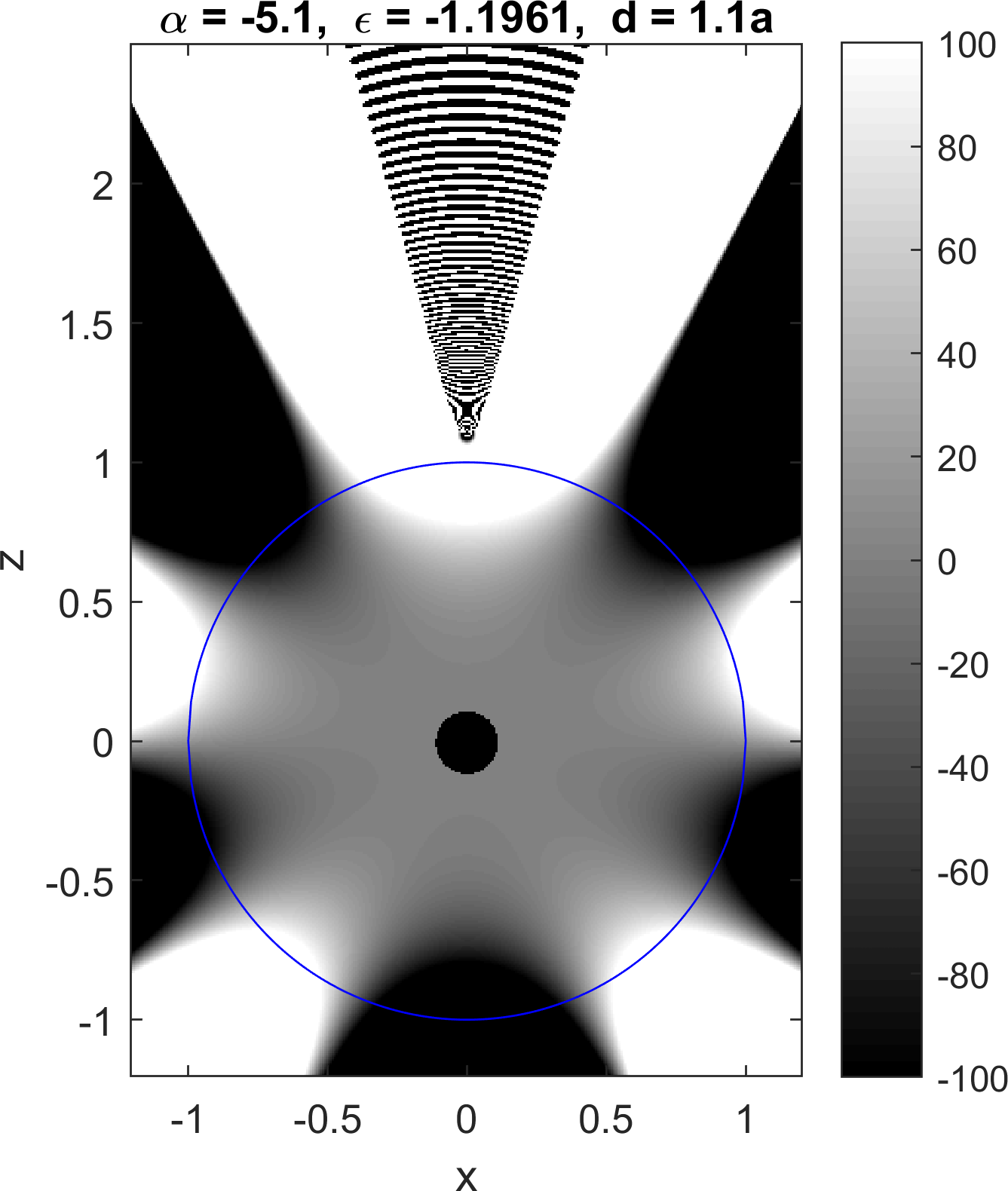}
(e)\includegraphics[scale=.451]{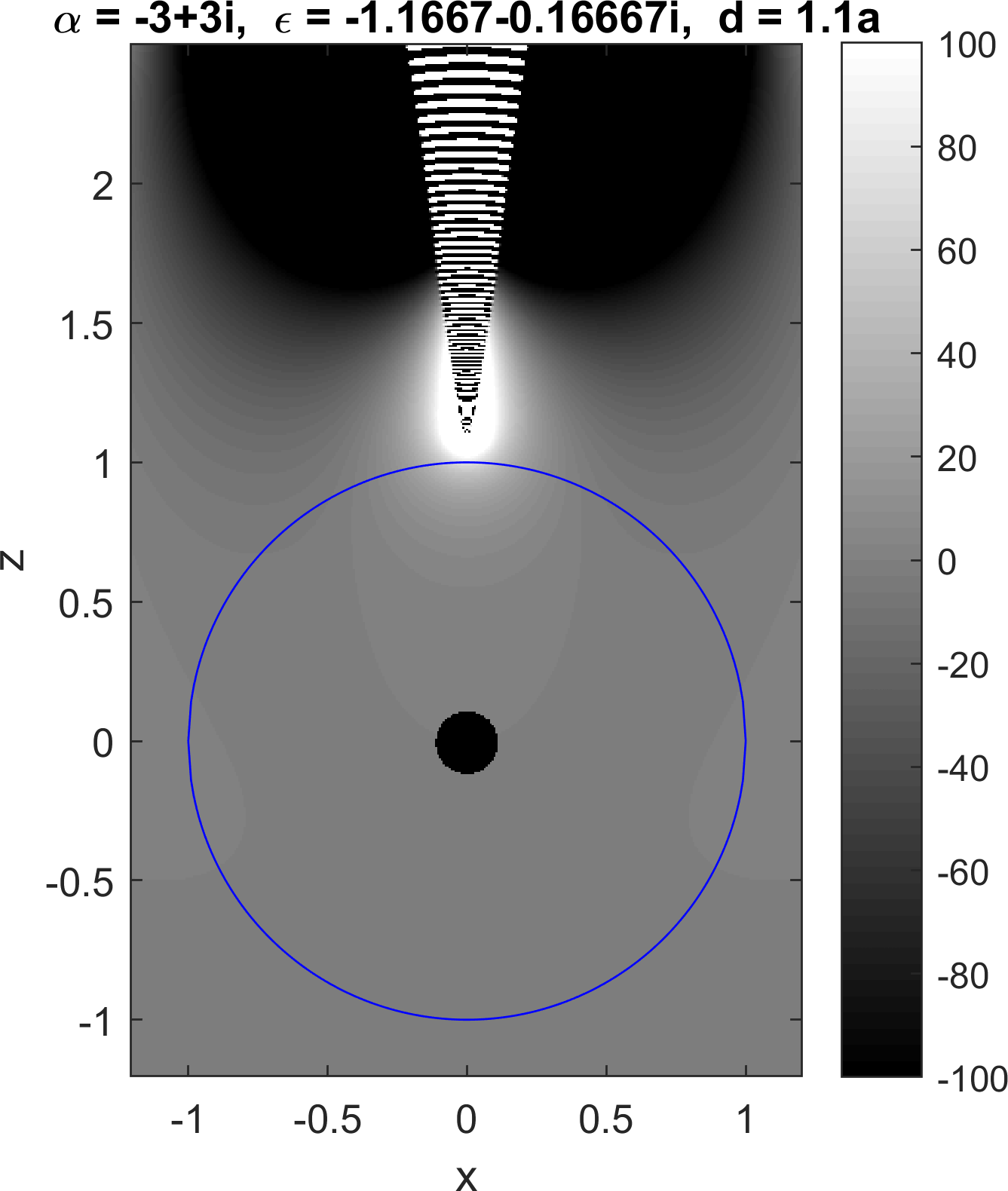}
(f)\includegraphics[scale=.451]{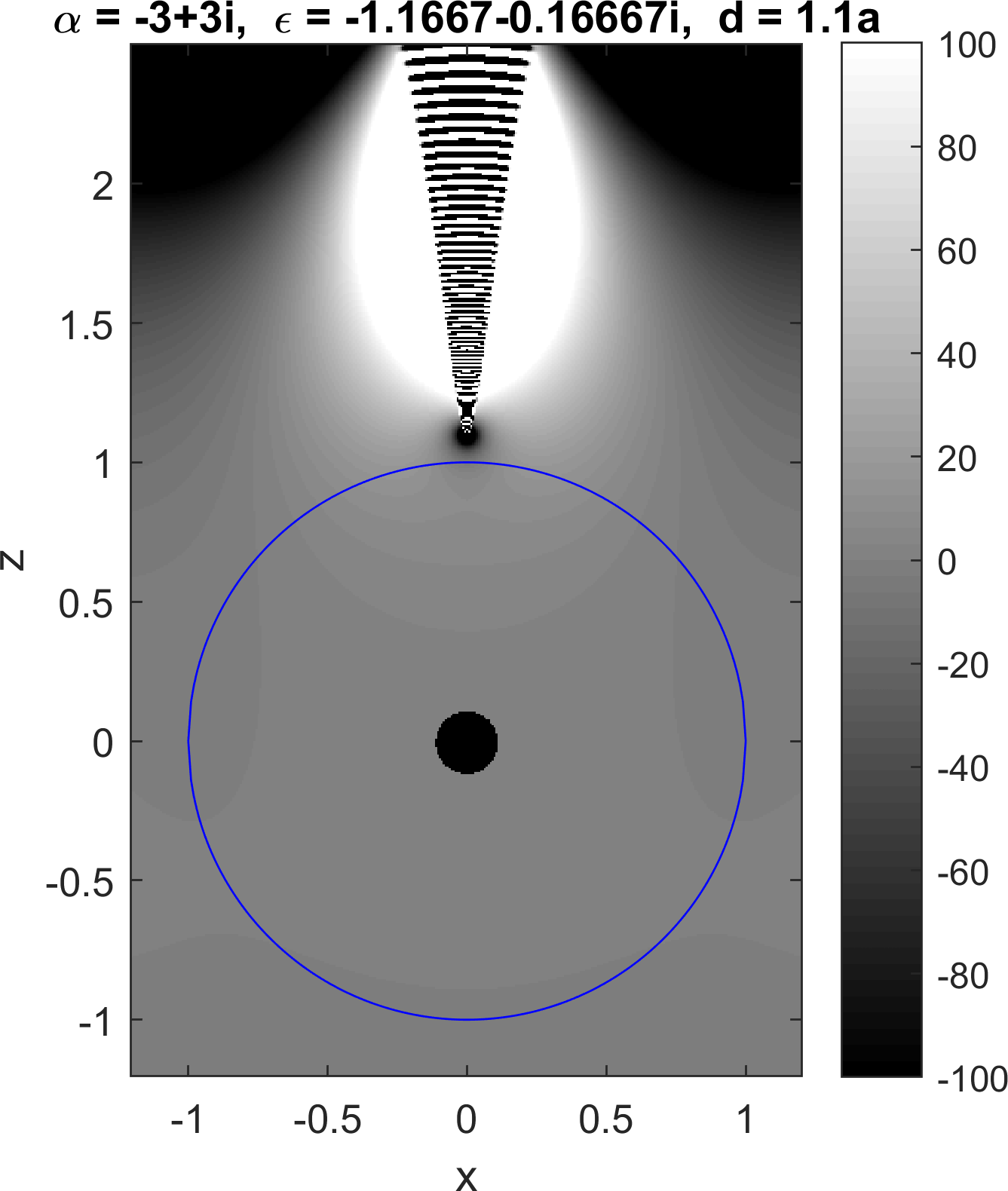}
\caption{Plots of the analytic continuation of the internal potential $V_i$ as computed with the inverted spheroidal harmonic series \eqref{VinQP}. The black circle near the origin is an artifact of the backwards recurrence used to compute the Legendre functions up to high orders; a truncation order of $n=300$ was chosen. The striped areas should also be ignored as they are an artifact from the slow convergence of the series near the image line. Note that (a),(b) have a different intensity scale. (e),(f) are real and imaginary parts for a single complex $\epsilon$. } \label{plots in}
\end{figure}

\subsection{Solution inside the sphere}
Here the image is an integral over the semi-infinite line segment extending from the source, and now the integral diverges at $r=\infty$ for $\alpha<1 ~(\epsilon<0)$. We present the final results for the potential as a spherical harmonic series, a regularised image system, and a spheroidal harmonic series \cite{majic2017inside}:
\begin{align}
V_i&=\frac{V_0a}{\epsilon+1}\bigg[\frac{2}{r_d} + \frac{1}{d}\sum_{n=0}^\infty\frac{\epsilon-1}{n(\epsilon+1)+1}\bigg(\frac{r}{d}\bigg)^n P_n(\cos\theta) \bigg]\\
&=\frac{V_0a}{\epsilon+1}\bigg[\frac{2}{r_d} + \frac{\epsilon-1}{(\epsilon+1)^2} \lim_{\nu\rightarrow\infty}\bigg(\int_1^\nu \frac{u^{-\alpha} \d u}{\sqrt{\rho^2+(z-du)^2}} - \frac{a}{d}\sum_{n=0}^{\lfloor-\alpha\rfloor+1} \frac{\nu^{-n-\alpha}}{n+\alpha} \bigg(\frac{r}{d}\bigg)^n P_n(\cos\theta)\bigg)\bigg] \\
&=\frac{V_0a}{\epsilon+1}\bigg[\frac{2}{r_d} + (\epsilon-1)\frac{2}{r}\sum_{n=0}^\infty(2n+1)\prod_{k=1}^n\frac{\alpha-k}{\alpha+k}~Q_n(\check\xi)P_n(\check\eta) \bigg],\label{VinQP}
\end{align}
where $\check\xi=\bar\xi(r\rightarrow a^2/r),~~ \check\eta=\bar\eta(r\rightarrow a^2/r)$ are radially inverted offset spheroidal coordinates. This time the regularisation is a sum of infinite magnitude \textit{external} multipoles. The spheroidal series \eqref{VinQP} is plotted for a range of complex $\epsilon$ in figure \ref{plots in}. 
%For $\epsilon$ near the poles, we see that $V_{in}$ may be come extremely large inside the sphere.

%\subsection{limiting cases}
%Here we check that the limiting cases in analysed in\cite{lindell1992} still hold for $\epsilon'<-1$.
%A. $d\rightarrow\infty$. 
%\begin{align}
%\b p = \frac{\epsilon-1}{\epsilon+1} \int_0^b 
%\end{align}

\section{dipole sources}\label{dipoles}
We now present the results for dipole sources. The main difference is an additional image point dipole. For perpendicular orientation:
\begin{align}
V_{\perp e}=& V_p\frac{a^2}{r_d^2}\cos\theta_d, \\
V_{\perp s}=& V_p \frac{a}{d}\frac{\epsilon-1}{\epsilon+1}\bigg[\frac{b^2}{r_b^2}\cos\theta_b + \frac{\epsilon}{\epsilon+1}\frac{b}{r_b} - \frac{\epsilon}{\epsilon+1} \sum_{n=0}^\infty \frac{1}{n(\epsilon+1)+1}\bigg(\frac{b}{r}\bigg)^{n+1} P_n(\cos\theta)\bigg] \\
=& V_p \frac{a}{d}\frac{\epsilon-1}{\epsilon+1}\Bigg[\frac{b^2}{r_b^2}\cos\theta_b + \frac{\epsilon}{\epsilon+1}\frac{b}{r_b}  
-\frac{\epsilon b}{(\epsilon+1)^2} \lim_{\nu\rightarrow0} \bigg(\int_\nu^1 \frac{u^{\alpha-1}\d u}{\sqrt{\rho^2+(z-bu)^2}} + \sum_{n=0}^{\lfloor-\alpha\rfloor} \frac{\nu^{n+\alpha}}{n+\alpha}\bigg(\frac{b}{r}\bigg)^{n+1} P_n(\cos\theta)\bigg)\Bigg]\\
=&V_p \frac{a}{d}\frac{\epsilon-1}{\epsilon+1}\bigg[\frac{b^2}{r_b^2}\cos\theta_b + \frac{\epsilon}{\epsilon+1}\frac{b}{r_b} - \frac{2\epsilon}{\epsilon+1} \sum_{n=0}^\infty(2n+1)\prod_{k=1}^n\frac{\alpha-k}{\alpha+k}~Q_n(\bar\xi)P_n(\bar\eta) \bigg],
\end{align}
where $V_p=p/(4\pi\epsilon_ma^2)$, $p$ is the source dipole moment, and $\theta_d$, $\theta_b$ are the colatitudes from the source point and inversion point respectively. 
As for the point charge, the infinite multipoles are added only for $\alpha'<0$ ($\epsilon'<-1$). For tangential orientation:
\begin{align}
V_{|| e}=& V_p\frac{a^2}{r_d^2}\sin\theta_d\cos\phi, \\
V_{|| s}=& V_p \frac{a}{d}\cos\phi\frac{\epsilon-1}{\epsilon+1}\bigg[-\frac{b^2}{r_b^2}\sin\theta_b  +  \sum_{n=1}^\infty \frac{1}{n(\epsilon+1)+1}\bigg(\frac{b}{r}\bigg)^{n+1} P_n^1(\cos\theta)\bigg] \\
=& V_p \frac{a}{d}\cos\phi\frac{\epsilon-1}{\epsilon+1}\Bigg[-\frac{b^2}{r_b^2}\sin\theta_b + \frac{b^2\rho}{\epsilon+1} \lim_{\nu\rightarrow0} \bigg(\int_\nu^1 \frac{u^\alpha\d u}{(\rho^2+(z-bu)^2)^{3/2}} + \sum_{n=1}^{\lfloor-\alpha\rfloor} \frac{\nu^{n+\alpha}}{n+\alpha}\bigg(\frac{b}{r}\bigg)^{n+1} P_n^1(\cos\theta)\bigg)\Bigg]
\\
=&V_p \frac{a}{d}\cos\phi\frac{\epsilon-1}{\epsilon+1}\bigg[-\frac{b^2}{r_b^2}\sin\theta_b + \frac{2}{\epsilon+1} \sum_{n=1}^\infty\frac{2n+1}{n(n+1)}\bigg(\prod_{k=1}^n\frac{\alpha-k}{\alpha+k}-1\bigg)Q_n^1(\bar\xi)P_n^1(\bar\eta) \bigg].
\end{align}
The image expressions for the potential of dipole sources are also derived in \cite{zurita2009quasi}, (restricted to $\alpha'>0$ and $\alpha'>-1$ for perpendicular and parallel dipoles respectively). The condition $\alpha'>-1$ is equivalent to $\epsilon$ lying outside a circle radius 1/2 centered at -3/2+0i on the complex plane.

\section{Non-spherical scatterers}
The analysis above only considers the sphere, since it is the only shape where the scattered potential has a known analytic continuation which also shows this problem of divergence. For comparison we compare a few other simple geometries.

The corresponding problem in 2d involving a uniform line of charge (point source in 2d) parallel to an infinite dielectric cylinder (circular disk in 2d) is solved simply with two image line sources inside the cylinder, and there are no problems of divergence except at $\epsilon=-1$ exactly. 

The same problem for the dielectric elliptic cylinder \cite{sten1996focal} involves an image on the focal strip, plus, if the source is close enough, an image line, and this solution again diverges only at $\epsilon=-1$. 

The problem of a point charge near a prolate (oblate) spheroid has an analytic solution which, if the source is not too close, can be interpreted as the potential of a charge distribution on the focal line (disk). Here the charge distribution is a series of spheroidal harmonics that converges again except for $\epsilon=-1$. This is curious since the spheroid is a generalisation of a sphere, an we would naturally expect the same divergence problem for $\epsilon<-1$. A rough intuition for this is that the line singularity of the scattered potential extends to both sides of the centre of the spheroid (unlike for the sphere), and this extended support is enough to accommodate the solution without the need for a singular charge distribution. If the point charge is too close to the spheroid, the series solution then diverges outside some inner spheroid, and the concept of an image charge distribution on a focal line (or disk) breaks down. For a close point charge on the axis of a prolate spheroid, the charge distribution has been re-expressed in a way to find faster convergence \cite{lindell2001dielectric}, but still does not find the fully reduced analytic continuation with which this analysis works from.

\section{Conclusion}
The image system for a point source near a dielectric sphere for negative values of permittivity necessitates the addition of regularisations in the form of infinite-magnitude multipoles. The spheroidal harmonic expansions however apply for all values of $\epsilon$ and reaffirm the existence of the additional regularisations. The image system changes drastically as $\epsilon$ crosses the poles $\epsilon_\infty=-2,-1.5,-1.33...$. Permittivity is always real and positive in electrostatics, but not necessarily in a quasistatic analysis, which applies for example to nano particles excited at optical frequencies. 
 
\section*{Acknowledgements}
This research was funded by a Victoria University of Wellington doctoral scholarship.

\bibliographystyle{elsarticle-num}
\bibliography{../libraryH}
\end{document}